\documentclass[lettersize,journal]{IEEEtran}
\usepackage{amsmath,amsfonts}
\usepackage{algorithmic}
\usepackage{algorithm}
\usepackage{array}
\usepackage{orcidlink}
\let\oldorcidlink\orcidlink
\renewcommand{\orcidlink}[1]{\hspace{0.2em}\raisebox{-0.28em}{\scalebox{1.5}{\oldorcidlink{#1}}}}
\usepackage{textcomp}
\usepackage{stfloats}
\usepackage{url}
\usepackage{verbatim}
\usepackage{graphicx}
\usepackage{cite}
\usepackage{multirow}
\usepackage{amssymb}
\usepackage{pifont}
\usepackage[table]{xcolor}
\usepackage{colortbl}
\usepackage{makecell}
\usepackage{threeparttable}
\usepackage{booktabs}
\usepackage{booktabs,array}
\usepackage{subcaption}
\usepackage[labelsep=period]{caption}
\hypersetup{hidelinks=true}

\newcommand{\xmark}{\ding{55}}
\newcommand{\high}{\ding{108}}
\newcommand{\low}{\ding{109}} 

\begin{document}

\title{\fontsize{19}{28}\selectfont P\textsuperscript{3}CDA: Privacy-Preserving and Provably Secure Cross Domain Authentication Scheme for Internet of Drones}

\author{\
  {\text{Chengqi Hou}\textsuperscript{\orcidlink{0000-0002-5961-4408}},~\IEEEmembership{Student Member,~IEEE}}, and {\text{Beibei Li}\textsuperscript{\orcidlink{0000-0002-0485-1975}}},~\IEEEmembership{Senior Member,~IEEE},
\\
  \text{Ziqing Zhu}\textsuperscript{\orcidlink{0000-0003-4868-0343}}, Yang You, and {\text{Licheng Wang}\textsuperscript{\orcidlink{0000-0001-8418-1897}}},~\IEEEmembership{Member,~IEEE}

  \vspace{-0.08cm}  

\thanks{Chengqi Hou, Beibei Li, and Ziqing Zhu are with the School of Cyber Science and Engineering, Sichuan University, Chengdu 610000, China. Ziqing Zhu is also with the Tianfu Jiangxi Laboratory, Chengdu 641419, China (e-mail: houchengqi@stu.scu.edu.cn; libeibei@scu.edu.cn; zhuziqing@stu.scu.edu.cn).}

\thanks{Yang You is with the NSFOCUS Technologies Group Co., Ltd., Beijing 100089, China (e-mail: youyang@nsfocus.com).}

\thanks{Licheng Wang is with the School of Cyberspace Science and Technology, Beijing Institute of Technology, Beijing 100081, China (e-mail: lcwang@bit.edu.cn).}}

\markboth{Journal of \LaTeX\ Class Files,~Vol.~14, No.~8, August~2021}%
{Shell \MakeLowercase{\textit{et al.}}: A Sample Article Using IEEEtran.cls for IEEE Journals}

\maketitle

\begin{abstract}
With the rapid expansion of the Internet of Drones (IoD) and the high mobility of drones, cross-domain interactions across geographically distributed domains are inevitable. As a critical foundation for IoD security, the design of cross-domain authentication schemes is becoming increasingly urgent. However, existing authentication schemes usually fail to balance security, efficiency, and identity privacy, rendering them inadequate for the stringent demands of IoD scenarios. To fill this gap, we propose P\textsuperscript{3}CDA, a \underline{P}rivacy-\underline{P}reserving and \underline{P}rovably Secure \underline{C}ross-\underline{D}omain \underline{A}uthentication scheme. First, we design an efficient pseudonym management mechanism that supports adaptive pseudonym generation and batch operations for registration, verification, and revocation. Second, we propose a structurally enhanced Merkle Hash Tree (MHT) that enables batch pseudonym updates, thereby preventing drones from storing excessive pseudonym data blocks. On this basis, we develop a cryptographic accumulator-based cross-domain authentication scheme that achieves anonymous authentication with authorized pseudonyms, while maintaining traceability and revocation of malicious drones. We perform a rigorous security analysis of P\textsuperscript{3}CDA and formally prove its security under the Canetti–Krawczyk (CK) adversary model. Extensive experiments further demonstrate that P\textsuperscript{3}CDA outperforms the state-of-the-art (SOTA) schemes in terms of computational, communication, and storage overhead. 
\end{abstract}


\section{Introduction}
\IEEEPARstart{T}{he} Internet of Drones (IoD) integrates drone technology with the Internet of Things (IoT) to create a highly flexible, scalable communication infrastructure. Characterized by low cost, high mobility, and strong autonomy, drones enable IoD deployment in challenging environments where traditional IoT is limited, including military operations, disaster relief, and environmental monitoring \cite{b21,b22}. In an IoD system, the airspace is divided into multiple domains based on geographic location. Each domain is managed by one or more Ground Control Stations (GCSs), which assist drones in environmental perception and real-time data transmission to the nearest station. In addition, GCSs collect sensory data and coordinate drone operations within the IoD domain via wireless communication technologies \cite{b23}, facilitating remote control and intelligent decision-making.

Due to their high mobility, drones often need to traverse multiple IoD domains during flight missions, leading to frequent cross-domain interactions. However, this also poses serious security and privacy challenges for existing IoD systems \cite{b24,b29}. Drones collect vast amounts of sensitive data, which attackers can exploit through eavesdropping or other cyber threats. Moreover, hijacked drones could be repurposed to transport hazardous materials, posing a significant risk to public safety \cite{b25,b26}. Additionally, given that drones often operate in sensitive areas, such as military zones, attackers could compromise device identities, track flight trajectories, and infer mission details, leading to severe privacy breaches.

\subsection{Research Gap}

To address unauthorized access and privacy leakage in IoD, a secure and privacy-preserving cross-domain authentication scheme is essential \cite{b27,b28}. A key requirement is conditional privacy protection of device identities, i.e., drones must be able to authenticate anonymously, while system administrators should retain the capability to trace the identities of malicious drones and revoke their authorization if required. There are two primary categories of potential solutions. The first is anonymous credentials \cite{b2,b3,b5,b7,b8,b37}, typically constructed using aggregate signatures, group  signatures, or zero-knowledge proofs. Although these schemes ensure strong privacy protection, their limited traceability hinders administrators from recovering the true identities of malicious drones. Additionally, anonymous credentials often require extensive bilinear pairing operations for verification, incurring significant computational and communication overhead for drones and making them impractical in resource-constrained IoD environments.

Another approach is the pseudonym technique \cite{b1,b34,b32,b36,b4}, where a drone obtains a long-term identity and subsequently requests short-term pseudonyms for authentication. Compared to anonymous credentials, pseudonym-based schemes are more efficient in terms of computation and communication, making them suitable for resource-constrained scenarios. However, this approach also faces several challenges. Using a single pseudonym for an extended period fails to ensure sufficient anonymity. To enhance privacy, drones may maintain multiple active pseudonyms concurrently, resulting in high storage overhead. However, existing solutions lack an efficient mechanism for revoking all pseudonym authorizations associated with malicious devices. Additionally, most existing schemes rely on a server for centralized pseudonym distribution, which incurs high administrative costs and raises security concerns. If the server is compromised, the pseudonyms and key materials of all drones may be exposed, thereby compromising the authentication process.

The above challenges give rise to the following research question: \textit{Can we design a conditional privacy-preserving cross-domain authentication scheme for IoD that minimizes computational, storage, and communication overhead, while enabling efficient tracking and revocation of malicious drones?}

\subsection{Our Contribution}

In this paper, we respond to these challenges by proposing P\textsuperscript{3}CDA, a cross-domain authentication scheme that features a novel pseudonym management mechanism. \textit{The first challenge lies in overcoming the performance bottlenecks and key exposure inherent in traditional centralized pseudonym assignment methods.} Therefore, a pseudonym Merkle Hash Tree (MHT) structure is designed in P\textsuperscript{3}CDA, enabling drones to autonomously generate batch pseudonyms and construct the corresponding MHT. By operating on fixed-size Merkle roots, drones can efficiently perform batch registration and verification of pseudonym data blocks, eliminating the need for a trusted third party to maintain key information. \textit{Unfortunately, maintaining anonymity and unlinkability still requires drones to store a large number of pseudonyms.} To tackle this issue, we enhance the pseudonym MHT using chameleon hash and integrate an adaptive pseudonym update approach into P\textsuperscript{3}CDA, substantially reducing the resource overhead associated with pseudonym updates. Drones can dynamically adjust the number of pseudonyms maintained in each time period according to available resources. When needed, drones perform batch updates to efficiently refresh all stored pseudonyms.

Building on this foundation, we divide P\textsuperscript{3}CDA into two core processes: intra-domain login and cross-domain authentication. Drones first obtain cross-domain authorization for their pseudonyms from the initial domain, enabling them to perform secure mutual authentication with GCSs in other domains. \textit{A critical challenge is ensuring the correctness and authenticity of the pseudonyms provided by drones, especially in the presence of malicious devices.} To address this, we employ Identity-Based Cryptography (IBC) to bind a drone’s identity to its Merkle root, treating it as public key-generated information. Furthermore, we implement traceability by embedding relevant parameters into pseudonym authorization tokens, enabling administrators to identify and revoke malicious drones when necessary. Instead of relying on Certificate Revocation Lists (CRLs), we utilize a cryptographic accumulator to ensure revocable cross-domain authorization, which enables efficient pseudonym revocation while maintaining authentication efficiency. Our contributions are summarized as follows:

\begin{itemize}
\item{We propose a pseudonym management mechanism that enables batch registration, verification, and updating of pseudonyms for drones. By integrating MHT, chameleon hash, and IBC, our mechanism overcomes the security and efficiency limitations of existing methods.}

\item{We propose a secure cross-domain authentication scheme for IoD scenarios that ensures anonymity, unlinkability, and revocability while enabling malicious drone revocation without relying on CRLs.}

\item{We conduct an in-depth security analysis of P\textsuperscript{3}CDA, provide a formal proof under the Canetti-Krawczyk (CK) model using the sequence-of-games approach, and perform automated security verification using the widely recognized tool ProVerif.}

\item{We simulate an IoD experimental environment to conduct a comprehensive performance evaluation. Comparisons with the existing SOTA schemes across computational, communication, and storage overheads demonstrate the efficiency and practicality of P\textsuperscript{3}CDA.}

\end{itemize}

\section{Related Work}
Several studies have already been conducted to achieve secure authentication across different trust domains. Tian \emph{et al.} \cite{b33} proposed a cross-domain authentication protocol for Unmanned Aerial Vehicles (UAV) that utilizes Physical Unclonable Functions (PUF) to enable secure communication across different domains. However, the scheme does not provide identity privacy protection during the authentication process. Wang \emph{et al.} \cite{b35} modeled authentication relationships as an undirected graph, and further proposed a secure cross-domain authentication scheme for IoT based on cryptographic accumulators and transferable signatures. Furthermore, Dong \emph{et al.} \cite{b6} designed a cross-domain authentication scheme for industrial networks by integrating certificateless cryptography with blockchain, and verified its security using SVO logic. However, this scheme also does not provide anonymity. 

\subsection{Pseudonym Technique based Authentication Schemes}

To enhance identity privacy, researchers have incorporated the pseudonym technique into cross-domain authentication, which are collectively classified as Pseudonym Technique-Based (PTB) schemes. Shen \emph{et al.} \cite{b1} proposed BASA, an anonymous cross-domain authentication scheme for multi-domain Industrial Internet of Things (IIoT). The scheme establishes trust among IIoT domains using a federated blockchain and integrates IBC to facilitate cross-domain authentication. However, BASA requires frequent interactions with the blockchain during authentication, leading to high latency in practical deployments. Yang \emph{et al.} \cite{b34} proposed a cross-domain authentication scheme for Internet of Vehicles (IoV), which relies on a Certificate Authority (CA) for centralized pseudonym distribution and employs blockchain to achieve secure cross-domain data interactions. However, the batch pseudonym distribution and update process introduce considerable communication overhead, and the scheme’s security and privacy are vulnerable to attacks targeting the CA. Lin \emph{et al.} \cite{b32} proposed a cross-domain authentication and key management scheme for fog computing based on Elliptic Curve Cryptography (ECC), offering high computational efficiency. However, this scheme fails to defend against impersonation and man-in-the-middle attacks. Khan \emph{et al.} \cite{b27} proposed a privacy-preserving authentication scheme for UAV-enabled intelligent transportation systems, leveraging Hyper Elliptic Curve Cryptography (HECC), digital signatures, and hash functions to enhance security and privacy. Miao \emph{et al.} \cite{b36} proposed an efficient authentication protocol for UAV-assisted IoV systems based on the ECC algorithms, and formally analyzed its security using Burrows-Abadi-Needham (BAN) logic. However, the protocol lacks adequate protection for session keys. Tian \emph{et al.} \cite{b4} designed a flexible authentication scheme based on the PUF for secure communication between IoT devices and UAVs, and formally validated its security using Mao-Boyd logic. However, this scheme relies on extra hardware support.

\begin{table*}[t]
\centering
\begin{threeparttable}
\caption{Functional Comparison of P\textsuperscript{3}CDA with Existing Works}
\renewcommand{\arraystretch}{1.2}
\setlength{\tabcolsep}{6.5pt}
\resizebox{0.95\textwidth}{!}{%
  \begin{minipage}{\textwidth}
    \begin{tabular}{c|cccc|cccc|c|cc}
      \specialrule{1.2pt}{1.2pt}{1.2pt}
      \multirow{2}{*}{Schemes} & 
      \multicolumn{4}{c|}{Security Strength} & 
      \multicolumn{4}{c|}{Privacy Strength} & 
      \multirow{2}{*}{Techniques} & 
      \multicolumn{2}{c}{Efficiency} \\
      \cmidrule(lr){2-5} \cmidrule(lr){6-9} \cmidrule(lr){11-12}
      & SMA & SKS & MDR & FSP & Anon. & Trace. & Unlink. & Type & & Comp. & Comm. \\
      \specialrule{1.2pt}{1.2pt}{1.2pt}

      TCD-DAA \cite{b35} & \checkmark & \xmark & \xmark & \xmark & \xmark & \xmark & \xmark & N/A & TS, Acc & \low & \high \\
      BCCA \cite{b6} & \checkmark & \checkmark & \xmark & \checkmark (DY) & \xmark & \xmark & \xmark & N/A & IBC & \high & \high \\
      BASA \cite{b1} & \checkmark & \checkmark & \xmark & \xmark & \checkmark & \xmark & \xmark & PTB & BC, IBC & \low & \low \\
      AS-CDA \cite{b32} & \xmark & \checkmark & \xmark & \checkmark (DY) & \checkmark & \xmark & \xmark & PTB & ECC, Hash & \high & \high \\
      AUAA \cite{b36} & \checkmark & \xmark & \xmark & \checkmark (BR) & \checkmark & \checkmark & \xmark & PTB & ECC, Hash & \high & \low \\
      PPS-UAA \cite{b4} & \checkmark & \checkmark & \xmark & \checkmark (DY) & \checkmark & \xmark & \checkmark & PTB & PUF, Hash & \high & \low \\
      XAuth \cite{b2} & \checkmark & \xmark & \xmark & \xmark & \checkmark & \xmark & \checkmark & ACB & BC, NIZK & \low & \low \\
      CCAP \cite{b3} & \checkmark & \xmark & \checkmark & \xmark & \checkmark & \checkmark & \checkmark & ACB & BC, NIZK & \low & \low \\
      BAR-CDA \cite{b8} & \checkmark & \xmark & \checkmark & \xmark & \checkmark & \checkmark & \checkmark & ACB & BC, GS & \low & \low \\
      SCDA \cite{b37} & \checkmark & \xmark & \checkmark & \xmark & \checkmark & \xmark & \checkmark & ACB & NIZK, Acc, IC & \high & \low \\
      \rowcolor[gray]{0.9}
      \textbf{P\textsuperscript{3}CDA}  & \checkmark & \checkmark & \checkmark & \checkmark (CK) &
      \checkmark & \checkmark & \checkmark & PTB & MHT, CH, Acc & \high & \high \\
      \specialrule{1.2pt}{1.2pt}{0pt}
    \end{tabular}
  \end{minipage}%
}
\end{threeparttable}

\vspace{0.5em}
\parbox{0.95\textwidth}{
\footnotesize
\textbullet\ \textbf{SMA} - Secure Mutual Authentication; \textbf{SKS} - Session Key Security; \textbf{MDR} - Malicious Device Revocation; \textbf{FSP} - Formal Security Proof; \textbf{Anon.} - Anonymity; \textbf{Trace.} - Traceability; \textbf{Unlink.} - Unlinkability; \textbf{N/A} - Not Applicable. \\
\textbullet\ \textbf{DY} - Dolev-Yao model; \textbf{BR} - Bellare-Rogaway model; \textbf{Acc} - Accumulator; \textbf{TS} - Transitive Signature; \textbf{BC} - Blockchain; \textbf{NIZK} - Non-Interactive Zero-Knowledge Proof; \textbf{GS} - Group Signature; \textbf{IC} - Integer Commitment; \textbf{CH} - Chameleon Hash.\\
\textbullet\ \checkmark\ indicates the scheme supports the function; \xmark\ indicates it does not; \ding{108} High efficiency; \ding{109} Low efficiency. \\
}
\end{table*}

\subsection{Anonymous Credential based Authentication Schemes}
In addition, recent studies have explored Anonymous Credential-Based (ACB) schemes, which leverage anonymous credentials to construct cross-domain authentication schemes that offer stronger privacy protection. Chen \emph{et al.} \cite{b2} proposed a privacy-preserving cross-domain authentication scheme called XAuth, in which devices achieve anonymous authentication using certificates combined with zero-knowledge proofs. However, XAuth lacks an effective revocation mechanism for handling malicious devices. To enable secure authentication across IoT domains with different cryptographic configurations, Tong \emph{et al.} \cite{b3} proposed a scheme called CCAP, which preserves the device's identity privacy using zero-knowledge proofs. Mood \emph{et al.} \cite{b5} proposed a secure anonymous authentication and key management scheme for multi-domain IoV, leveraging blockchain and dual signatures to reduce the number of message interactions. Zeng \emph{et al.} \cite{b7} proposed an anonymous cross-domain authentication scheme for IIoT that integrates dynamic group signatures with blockchain technology and supports revocation of malicious devices. However, the scheme incurs high computational overhead. To address the security and privacy requirements of cross-domain interactions in vehicular ad-hoc networks, Li \emph{et al.} \cite{b8} employed blockchain for vehicle identity management and proposed a cross-domain authentication scheme based on random short signatures. Chen \emph{et al.} \cite{b37} proposed an anonymous authentication scheme for multi-domain IIoT that combines RSA accumulators, zero-knowledge proofs, and integer commitment to generate anonymous credentials for devices. However, the scheme lacks an effective and traceable mechanism for malicious participants when necessary.

In Table I, we compare P\textsuperscript{3}CDA with three categories of SOTA cross-domain authentication schemes: non-anonymous, pseudonym technology-based and anonymous credential-based schemes. Overall, existing solutions fail to strike a balance between security, privacy, and efficiency. Although the non-anonymous schemes offer high efficiency, they do not provide protection of identity privacy. PTB schemes suffer from limited unlinkability and risk revealing associations across multiple authentication requests. ACB schemes typically rely on computationally expensive bilinear pairings, rendering them unsuitable for resource-constrained environments. Therefore, our research aims to fulfill the current research gap.	

\section{System and Adversary Models}
In this section, we describe the system model, adversary model, and security goals defined in P\textsuperscript{3}CDA.

\subsection{System Model}
As shown in Fig. 1, the IoD system consists of four types of entities, each with distinct responsibilities in the interaction process, as detailed below.

\begin{itemize}
\item{\textbf{Trusted Authority (TA):} The TA serves as the administrator of the IoD system, responsible for publishing public parameters, assisting other entities with key generation, and tracking the true identity of a drone when it exhibits malicious behaviours.}

\item{\textbf{Domain Enrollment Authority (DEA):} Each IoD domain is equipped with a DEA that assists the TA in managing that domain. The DEA receives intra-domain login requests from drones, verifies these requests, and grants cross-domain authorization for verified pseudonyms.}

\item{\textbf{Ground Control Station (GCS):} The GCS is distributed across IoD domains in different geographic locations. It is responsible for authenticating drones, establishing secure communication sessions, and obtaining sensory data from drones within its communication range. Additionally, the GCS handles flight task assignments and provides relevant services to drones.}

\item{\textbf{Drones:} Drones operate across multiple IoD domains, performing flight tasks while sensing relevant data and establishing communication connections with neighboring GCSs. This ensures secure data transmission and allows drones to access authorized services.}

\end{itemize}

\begin{figure}
\centerline{\includegraphics[width=0.49\textwidth]{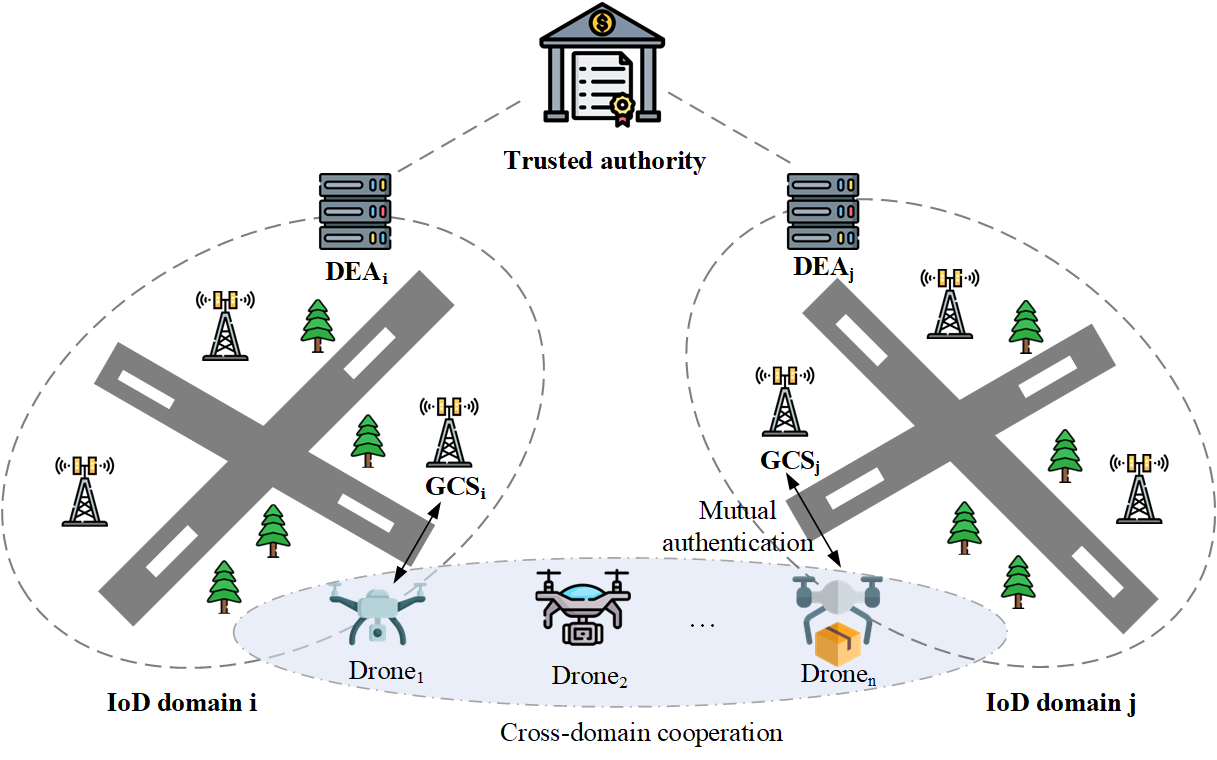}}
\caption{System model.}
\label{fig}
\end{figure}

\subsection{Adversary Model}

To construct a capable adversary model, we adopt the widely accepted Dolev-Yao (DY) \cite{b15} and Canetti-Krawczyk (CK) \cite{b11} models. In the DY model, the adversary can initiate an arbitrary number of session requests and fully control the transmissions over public channels between entities. Specifically, the adversary $\mathcal{A}$ is allowed to eavesdrop, replay, delete, tamper with, and forge messages over all public channels. Once the session key is compromised, $\mathcal{A}$ can decrypt messages and retrieve their contents. To simulate the potential risk posed by insider attackers, we extend the adversary's capabilities. Specifically, $\mathcal{A}$ is allowed to access all confidential information maintained by the TA, including the key materials distributed to each drone, while its private keys remain inaccessible.

In the formal security proof, we employ the stronger CK model that encompasses all adversarial capabilities defined in the DY model. To evaluate forward secrecy, we assume that $\mathcal{A}$ can access secret values stored on drones, including long-term keys and other secrets. This assumption is reasonable and practical given the limited computational and storage capabilities of drones. The detailed definition of the adversary model is discussed in Section VI.

Furthermore, in P\textsuperscript{3}CDA, the GCS is regarded as a semi-honest entity, which faithfully follows the scheme but may try to infer private information from its interactions with drones. In contrast, drones are treated as untrusted entities, some of which may behave maliciously. These malicious drones may collude with attackers to gain unauthorized access to IoD domains through various attack vectors. Malicious drones may engage in misbehavior across multiple domains and attempt to evade the traceability and accountability mechanisms enforced by the TA and ETA.

\subsection{Security Goals}
The scheme must be designed to meet the following goals.

\begin{itemize}
\item{\textbf{Mutual authentication:} Through message exchanges in the scheme, registered legitimate drones achieve secure mutual cross-domain authentication with GCSs across different IoD domains.}

\item{\textbf{Secure session key:} After successful authentication, drones can negotiate a secure session key with the GCS to ensure secure data communication.}

\item{\textbf{Conditional anonymity:} Drones are required to maintain anonymity in cross-domain interaction, preventing attackers from inferring their identities. In addition, TA collaborates with DEA to trace malicious drones and revoke both their identities and authorized pseudonyms, thereby enhancing the robustness of the IoD system.}

\item{\textbf{Unlinkability:} Drones' pseudonyms should be unlinkable, meaning that an attacker cannot determine whether multiple cross-domain authentication requests originate from the same device. }

\item{\textbf{Forward secrecy:} The scheme should ensure forward secrecy, meaning that even if a drone's long-term or short-term secrets are disclosed, the secrecy of its previous session keys remains unaffected.}

\item{\textbf{Common attacks resistance:} The scheme should be resistant to common security attacks, including man-in-the-middle, replay, privileged insider attacks, and other potential threats.}

\end{itemize}

\section{Preliminaries}
This section introduces the preliminaries of P\textsuperscript{3}CDA.

\subsection{Computational Assumptions}
\textbf{Elliptic Curve Computational Diffie-Hellman (ECCDH) Assumption.} \textit{Given $(\mathbb{G}, G, a \cdot G, b \cdot G)$, where $\mathbb{G}$ is an elliptic curve cyclic group, $G$ is a generator, and $a, b$ are randomly chosen on $\mathbb{F}_p$. The advantage for adversary $\mathcal{A}$ to compute $a \cdot b \cdot G$ in polynomial time $t$ is: $\mathit{Adv}^{\mathit{ECCDH}}(t) \leq \varepsilon_{cdh}$.}

\textbf{q-Strong Diffie-Hellman (q-SDH) Assumption \cite{b12}.} \textit{Given $(\mathbb{G}, G, s \cdot G, ...,{s^q} \cdot G)$, where $\mathbb{G}$ is an elliptic curve cyclic group, $G$ is a generator, and $s$ is randomly chosen on $\mathbb{F}_p$. The advantage for adversary $\mathcal{A}$ to compute a pair $\left(x, {\frac{1}{x+s}} \cdot G\right)$ in polynomial time $t$ is: $\mathit{Adv}^{\mathit{q-SDH}}(t) \leq \varepsilon_{sdh}$, where $x \in \mathbb{F}_p$.}

\subsection{Bilinear Pairing}
Bilinear pairing \cite{b41}, as a fundamental primitive in cryptography, has been widely used in numerous public key cryptographic protocols. In this paper, we adopt the Type-I bilinear pairing setting. Let $\mathbb{G}$ and $\mathbb{G}_T$ be two cyclic groups defined over a prime order $p$. A bilinear pairing is defined as a map $e: \mathbb{G} \times \mathbb{G} \to \mathbb{G}_T$ with the following properties.

\begin{itemize}
    \item \textbf{Bilinearity}: For all $a, b \in \mathbb{F}_p$, and $P, Q \in \mathbb{G}$, we can get 
    $e(a \cdot P, b \cdot Q) = e(P, Q)^{a \cdot b}$.
    
    \item \textbf{Non-degenerate}: $e(P, Q) \ne 1$.
    
    \item \textbf{Computability}: $e(P, Q)$ can be computed efficiently.
\end{itemize}

\subsection{Chameleon Hash}
The chameleon hash function \cite{b31}, introduced by Krawczyk and Rabin in 1998, is a trapdoor-based hash function. For participants without knowledge of the trapdoor, it behaves like a traditional hash function, maintaining fundamental security properties such as collision resistance. However, for entities holding the trapdoor, it enables efficient computation of hash collisions for arbitrary messages. A chameleon hash function consists of the following components.
\begin{itemize}

\item{\textbf{$\text{CH.KeyGen}(1^\lambda) \rightarrow (pk, sk)$}: Given the input security parameter $\lambda$, the algorithm outputs a public-private key pair $(pk, sk)$ for hash computation.}

\item{\textbf{$\text{CH.Hash}(pk, m) \rightarrow (h, r)$}: Given an input message $m$, the algorithm computes a hash value $h$ and a random number $r$ for this message using the public key $pk$.}

\item{\textbf{$\text{CH.Verify}(pk, m, h, r) \rightarrow \{0,1\}$}: Given an input message $m$ and the corresponding $(h, r)$, the algorithm verifies the correctness of the hash value using the public key $pk$, outputting 1 if the verification is successful and 0 otherwise.}

\item{\textbf{$\text{CH.Adapt}(sk, m, m', h, r) \rightarrow r'$}: Given an input message $m$ and the corresponding $(h, r)$, the algorithm generates a hash collision using the private key $sk$ to compute a new random number $r'$ for a new message $m'$.}

\end{itemize}

However, some classic chameleon hash functions are vulnerable to key-exposure attacks, allowing adversaries to generate hash collisions. To address this issue, several key-exposure-resistant chameleon hash functions have been proposed. In this paper, we adopt one of the efficient constructions \cite{b9} to enhance security and robustness against key exposure attacks.

\subsection{Cryptographic Accumulator}
Cryptographic accumulators offer an efficient method for managing element sets. Given an element set $S$, the accumulator computes a fixed-size cumulative value $v$ to represent $S$, and generates a witness $w$ for each element in the set. Using the element witness $w$, a verifier can efficiently confirm membership, determining whether a given element belongs to $S$. An accumulator is constructed as follows.

\begin{itemize}

\item{\textbf{$\text{Acc.Setup}(1^\lambda) \rightarrow (pk, sk)$}: Given the input security parameter $\lambda$, the algorithm sets the system parameters and outputs the public-private key pair $(pk, sk)$ for the accumulator.}

\item{\textbf{$\text{Acc.Eval}(X, sk) \rightarrow v$}: Given an input element set $X = \{x_1, x_2, \dots, x_t\}$, the algorithm computes a cumulative value $v$ representing this element set using the accumulator private key $sk$.}

\item{\textbf{$\text{Acc.Witness}(x_i, sk, v) \rightarrow w_i$}: Given an input element $x_i$ and a cumulative value $v$, the algorithm determines whether $x_i$ belongs to the element set $X$ and outputs a witness $w_i$ for the correct element.}

\item{\textbf{$\text{Acc.Verify}(x_i, w_i, v, pk) \rightarrow \{0,1\}$}: Given an input element $x_i$, the witness $w_i$, and the cumulative value $v$, the algorithm verifies its correctness using the accumulator public key $pk$ and outputs the result $\{0,1\}$.}

\end{itemize}

Since modifying the set, by adding or removing an element, causes a change in the cumulative value $v$, classical accumulators typically require updating the witnesses of remaining elements. Under the considered scenarios, this leads to significant computational and communication overhead. To address this issue, we adopt a more efficient dynamic accumulator construction \cite{b10} that enables an efficient element addition.

\begin{figure}
\centerline{\includegraphics[width=0.49\textwidth]{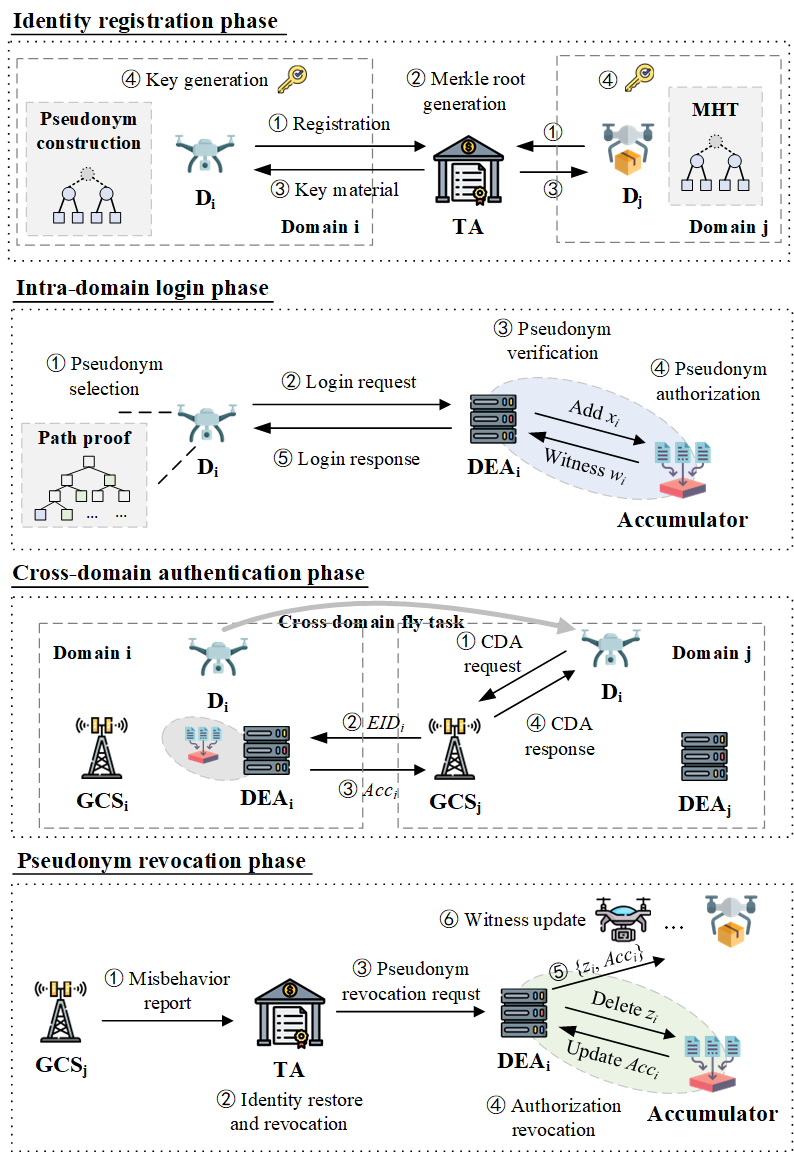}}
\caption{Overview of P\textsuperscript{3}CDA.}
\label{fig}
\end{figure}

\section{The Proposed Scheme}
In this section, we present a detailed construction of P\textsuperscript{3}CDA.

\subsection{Technical Overview}
As shown in Fig. 2, P\textsuperscript{3}CDA consists of five core phases: identity registration, intra-domain login, cross-domain authentication, pseudonym revocation, and pseudonym update. In the identity registration phase, the drone generates a batch of pseudonyms and constructs a pseudonym MHT. Using the IBC, the TA binds the drone’s public key to both its device identity and the Merkle root, enabling simultaneous registration of the identity and batch pseudonyms. In the intra-domain login phase, by presenting Merkle path proofs and public key, the drone can efficiently prove the authenticity and correctness of its pseudonym to the ETA. The ETA authorizes verified pseudonyms for cross-domain authentication using the accumulator. Due to the efficient accumulator construction adopted in this paper, adding new elements to the accumulator does not require updating the witnesses of other drones, ensuring a highly efficient process.

\begin{figure*}[htbp]
\centering
{\includegraphics[width=0.97\textwidth]{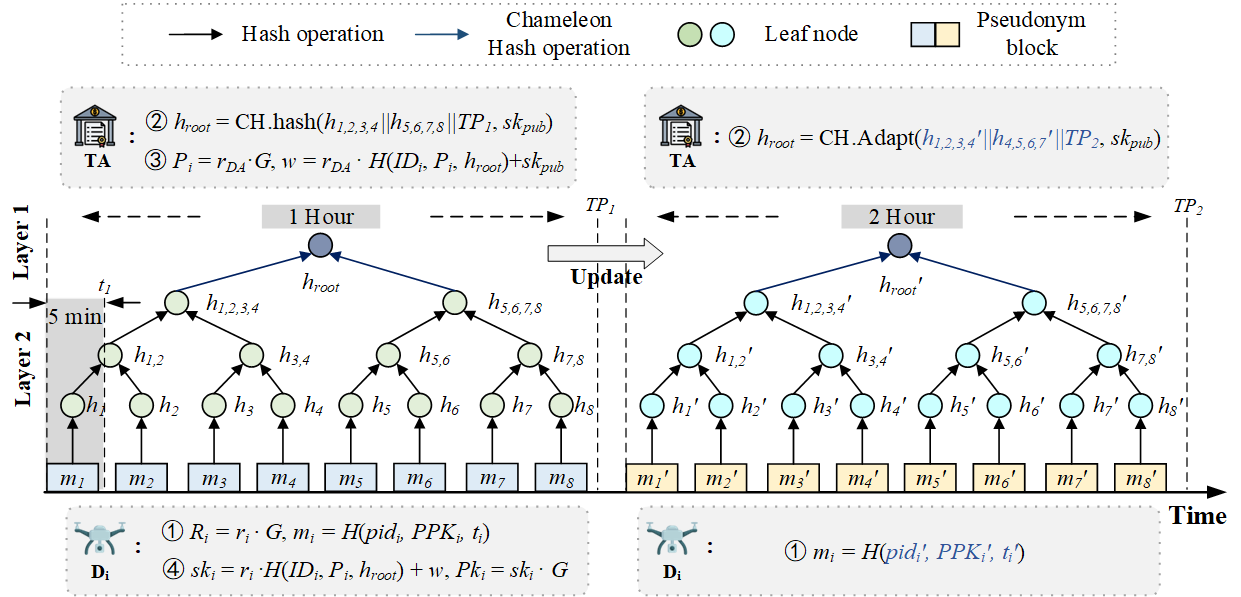}}
\caption{Example of registering and updating batch pseudonyms in P\textsuperscript{3}CDA.}
\label{fig}
\end{figure*}

In the cross-domain authentication phase, the drone uses its authorized pseudonym to perform mutual authentication with GCSs in different domains and negotiates a session key to ensure the security and reliability of subsequent data communications. In addition, we design a revocation phase that enables the TA to trace the identity of a malicious drone via its authorization token. The DEA can further remove the drone's pseudonym information from the element sets in the dynamic accumulator, achieving efficient pseudonym revocation without relying on a CRL. In the pseudonym update phase, the identity registration process must typically be re-executed to update the pseudonym data blocks, since the drone’s public key is bound to the Merkle root. To overcome this, we enhance the MHT structure using a chameleon hash function. Specifically, the TA manages the private key of the chameleon hash function, which is used to generate the root node of the drone’s pseudonym MHT. By leveraging controlled hash collisions, we enable an efficient pseudonym update process, thereby eliminating the need for frequent re-registration and effectively reducing resource consumption.

\subsection{Initialization Phase}
The initialization phase sets up the IoD system between the TA and the DEAs, involving the negotiation and distribution of public parameters.

\textbf{Step 1:} Given a security parameter $\lambda$, the TA first defines a finite field $\mathbb{F}_p$ with a prime order $p$, two elliptic curve cyclic group $\mathbb{G}$ and $\mathbb{G_T}$, two generators $G, G_1 \in \mathbb{G}$, and a bilinear pairing function $e: \mathbb{G} \times \mathbb{G} \to \mathbb{G_T}$. Additionally, the TA defines a secure hash function $H(\cdot)$. After the setup, the TA selects a random number $sk_{\text{pub}} \in \mathbb{F}_p$ as its private key and further computes its public key $PK_{\text{pub}} = sk_{\text{pub}} \cdot G$.

\textbf{Step 2:} $\text{DEA}_{i}$, as the administrator of IoD domain $i$, selects its private key $sk_{\text{ETA}}^i \in \mathbb{F}_p$ and computes the public key $PK_{\text{ETA}}^i = sk_{\text{ETA}}^i \cdot G$. Furthermore, $\text{DEA}_{i}$ initializes a cryptographic accumulator for domain $i$. It selects $y_i \in \mathbb{F}_p$ as the accumulator's private key and computes its public key $Y_i = y_i \cdot G_1$. To initialize the accumulator value, $\text{DEA}_{i}$ chooses a random number $
u_i \in \mathbb{F}_p$ and computes $Acc_i = u_i \cdot G_1$. After these computations, the TA and $\text{DEA}_{i}$ jointly release the system's public parameters $
\{G, G_1, H(\cdot), PK_{\text{pub}}, EID_i, PK_{\text{ETA}}^i, Y_i, Acc_i\}$, where $EID_i$ represents the identity of $\text{DEA}_{i}$.

\subsection{Identity Registration Phase}
The identity registration phase ensures secure enrollment of GCSs and drones through a secure channel. Drones generate a batch of pseudonym data blocks, organize them into a pseudonym MHT, and submit the resulting root together with their identities to the TA to derive their long-term keys. The flow of this registration process is illustrated in Fig. 3.

\textbf{Step 1:} A drone $\text{D}_{i}$ first selects a random number $r_i \in \mathbb{F}_p$ and computes $
R_i = r_i \cdot G$, which is used to request long-term keys from the TA. For pseudonym generation, $\text{D}_{i}$ sets a long time period $TP_i$ and determines the number of pseudonyms $N$ it will use within this period based on its resources and privacy needs. $\text{D}_{i}$ then generates a pseudonym set $\text{pid} = \{\text{pid}_1, \text{pid}_2, \dots, \text{pid}_N\}$ and a random number set $z = \{z_1, z_2, \dots, z_N\}$, where $z_k \in \mathbb{F}_p, k \in \{1, \dots, N\}$. For the $k$-th pseudonym, $\text{D}_{i}$ computes its associated temporary key $psk_k = H(\text{pid}_k, z_k)$, $PPK_k = {psk}_k \cdot G$. After these computations, the drone constructs a pseudonym MHT, using pseudonym information as data blocks, including $\{\text{pid}_i, PPK_i, t_i\}$, where $i$ denotes the pseudonym index, and $t_i$ represents the expiration time of this pseudonym, satisfying $i < j$, $t_i < t_j < TP_i$. $\text{D}_{i}$ then computes the leaf node $
L_k = H(\text{pid}_i, PPK_i, t_i)$ of the Merkle tree and generates all non-leaf nodes except the Merkle root. Lastly, $\text{D}_{i}$ submits the registration information to TA, including $({ID}_i, R_i, LR_1, LR_2, TP_i)$, where $LR_1$ and $LR_2$ are the two nodes in the second layer of the Merkle tree, and the root node is computed by hashing these two nodes. Additionally, the ${ID}_i$ represents the unique identifier for $\text{D}_{i}$.

\textbf{Step 2:} When the TA receives the registration request, it first verifies that ${ID}_i$ has not been registered before. If passed, TA proceeds to generate the root of the pseudonym MHT for $\text{D}_{i}$ using a chameleon hash. TA selects two random numbers $r_{\text{root}},k \in \mathbb{F}_p$ and sets the trapdoor key $tk = (r_{\text{root}}, sk_{\text{pub}})$. TA computes the Merkle root as $h_{\text{root}} = r_{\text{root}} \cdot G$, and the verification parameters $\delta = (r, K)$, $K = k \cdot G$, and $r = r_{\text{root}} - H(LR_1, LR_2, TP_i, K) \cdot (k + sk_{\text{pub}})$. Additionally, to assist $\text{D}_i$ in generating its long-term keys, TA selects $r_{\text{DA}} \in \mathbb{F}_p$, and computes $P_i = R_i + r_{\text{DA}} \cdot G$, $h_i = H({ID}_i, P_i, h_{\text{root}})$, and $w = h_i \cdot r_{\text{DA}} + sk_{\text{pub}}$. Once the computation is complete, TA returns $(P_i, w, \delta, h_{\text{root}})$ to $\text{D}_i$ and records this information in the registration list.

\textbf{Step 3:} After $\text{D}_i$ receives the registration response message, it first generates its long-term keys by computing $h_i' = H({ID}_i, P_i, h_{\text{root}})$, $sk_i = r_i \cdot h_i' + w$, and $PK_i = sk_i \cdot G$. $\text{D}_i$ verifies the correctness of the response message using $PK_i = P_i \cdot h_i' + PK_{\text{pub}}$. If  verified, $\text{D}_i$ completes the registration process. Otherwise, it re-executes this process.

\textbf{Remark.} The GCS registration process follows a similar procedure but does not require the pseudonym registration.

\subsection{Intra-Domain Login Phase}
To obtain cross-domain authorization for its pseudonym, the drone proceeds to the intra-domain login phase within its initial domain. The initial domain can be any IoD domain within the system and is considered a trusted domain for $\text{D}_{i}$, typically the one from which $\text{D}_{i}$ takes off for its mission. 

\textbf{Step 1:} $\text{D}_{i}$ selects a pseudonym with the closest expiration time, denoted as $\{\text{pid}_k, {psk}_k, {PPK}_k\}, k \in \{1, \dots, N\}$. $\text{D}_{i}$ then generates its Merkle path proof according to the pseudonym MHT, represented as $\pi = (L_1, L_2, \dots, L_s)$, $s = \lceil \log_2 N \rceil$. To maintain anonymity, $\text{D}_{i}$ conceals its identity by computing $S_1 = ({ID}_i || P_i) \oplus H(\text{pid}_k || {psk}_k \cdot PK_{\text{ETA}}^i)$. Since path proofs may contain repeated nodes in multiple requests, this could lead to linkability of pseudonym information. To prevent this, $\text{D}_{i}$ generates a random number $r_s$ and computes $S_2 = r_s \oplus H(P_i, S_1)$, $\pi^{*} = (L_{1}^{*}, L_{2}^{*}, \ldots, L_{s}^{*})$, $L_k^* = H_1(S_2, r_s) \oplus L_k$, $k \in \{1,s\}$. Lastly, $\text{D}_{i}$ records the current timestamp $T_1$ and computes $h_1 = H_1(\text{pid}_k, {PPK}_k, t_k, \pi^*, S_1, S_2, \delta, TP_i, T_1)$, $\sigma_1 = sk_i + h_1 \cdot {psk}_k$. After the computation, $\text{D}_{i}$ sends an intra-domain login request to $\text{DEA}_{i}$, including $(\text{pid}_k, {PPK}_k, t_k, \pi^*, S_1, S_2, \delta, TP_i, T_1, \sigma_1)$.

\textbf{Remark.} The authentication parameters $\delta$ and $TP_i$ of the chameleon hash are not hidden in this message, as $\text{D}_{i}$ is only required to provide these parameters in the first intra-domain login request. The DEA can then store these parameters and delete them when $TP_i$ expires, waiting for the drone’s next intra-domain login to update these parameters.

\textbf{Step 2:} Upon receiving the message, $\text{DEA}_{i}$ first records the current timestamp $T_1'$ and checks whether $T_1' - T_1 \leq \Delta T$, where $\Delta T$ is the maximum response time. Next, $\text{DEA}_{i}$ recovers $\text{D}_{i}$'s identity information ${ID}_i' || P_i' = S_1 \oplus H(\text{pid}_k || sk_{\text{ETA}}^i \cdot PPK_k)$. Additionally, $\text{DEA}_{i}$ reconstructs the Merkle path proof $\pi'$, by computing $r_s' = S_2 \oplus H(P_i', S_1)$, $L_k' = L_k^* \oplus H(S_2, r_s')$. Then $\text{DEA}_{i}$ obtains $LR_1'$ and $LR_2'$ from the Merkle path proof. To verify the correctness of the pseudonym data block $l_k' = H(\text{pid}_k, PPK_k, t_k)$, it checks $t_k < TP_i$. Using the validation parameter $\delta$, $\text{DEA}_{i}$ computes the chameleon hash $h_{\text{root}}' = H(LR_1', LR_2', TP_i, K) \cdot (K + PK_{\text{pub}}) + r' \cdot G$. Lastly, it verifies whether $\sigma_1 \cdot G = H({ID}_i', P_i', h_{\text{root}}') \cdot P_i + PK_{\text{pub}} + h_1' \cdot PPK_k$, where $h_1' = H(\text{pid}_k, PPK_k, t_k, \pi^*, S_1, S_2, \delta, TP_i, T_1)$.

\textbf{Step 3:} If the validation passes, $\text{DEA}_{i}$ authorizes this pseudonym information. Specifically, $\text{DEA}_{i}$ computes $V_i = {ID}_i' \oplus H(sk_{\text{ETA}}^i \cdot PK_{\text{pub}} + H(\text{pid}_i, PPK_i) \cdot G)$, and $x_i = H(\text{pid}_i, PPK_i, t_i, V_i)$. To achieve cross-domain authorization, $\text{DEA}_{i}$ add $x_i$ to the accumulator in domain $i$ and generates witness for it by computing $Acc_i' = Acc_i$, and $w_i = \frac{1}{y_i + x_i} \cdot Acc_i'$. The cross-domain authorization token for this pseudonym is obtained as $TK_i = (V_i, w_i)$. $\text{DEA}_{i}$ returns the response message, including $(TK_i, T_2, h_2)$, where $h_2 = H(TK_i, T_2, r_s')$.

\textbf{Step 4:} $\text{D}_{i}$ receives the response message and verifies the correctness of $h_2$ and $T_2$. If the verification is successful, it stores the cross-domain authorization token $TK_i$ for this pseudonym; otherwise, $\text{D}_{i}$ re-executes this phase.

\begin{figure}
\centerline{\includegraphics[width=0.49\textwidth]{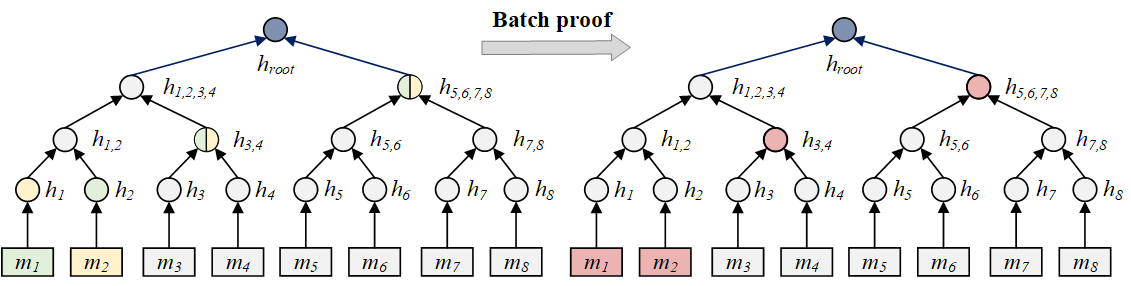}}
\caption{Path proof for data blocks $m_1$ and $m_2$ in the pseudonym MHT, with green denoting the path proof for $m_1$, yellow for $m_2$, and red for the batch proof for $m_1$ and $m_2$.}
\label{fig}
\end{figure}

\textbf{Batch pseudonym login:} When a drone undertakes a mission that requires traversing multiple IoD domains or operating for an extended period, it must submit multiple intra-domain login requests within the initial domain to maintain sufficient anonymity. To enhance efficiency, we introduce pseudonym batch logins, allowing drones to include multiple pseudonyms in a single login request for cross-domain authorization.

The drone first selects multiple pseudonyms with the closest expiration times, utilizing the design of the enhanced MHT to ensure that these pseudonym blocks are positioned closely together. As shown in Fig. 4, this structure results in Merkle path proofs containing a large number of repeated nodes, enabling a highly efficient path proof during batch intra-domain login \cite{b30}. Additionally, to maintain unlinkability, all pseudonyms in the request, except for the first pseudonym, must be concealed in the same way as the identity information $(ID_i, P_i)$. The signature of the request message can also be verified in batch to improve efficiency. For a set of pseudonym information $\{{(psk}_1, {PPK}_1), ({psk}_2, {PPK}_2), \dots, ({psk}_k, {PPK}_k)\}$, the drone computes $\sigma_1 = sk_i + h_1 \cdot \sum_{j=1}^{k} {psk}_j$. $\text{DEA}_{i}$ can then batch verify the correctness of the pseudonyms using $\sigma_i \cdot G = P_i \cdot H_1({ID}_i', P_i', h_{\text{root}}') + PK_{\text{pub}} + h_1' \cdot \sum_{j=1}^{k} \text{PPK}_j$.

\subsection{Cross-Domain Authentication Phase}
After completing intra-domain login, drones can use their authorized pseudonyms to perform anonymous cross-domain authentication with GCSs in other domains. The main process is illustrated in Fig. 5.

\begin{figure}
\centerline{\includegraphics[width=0.47\textwidth]{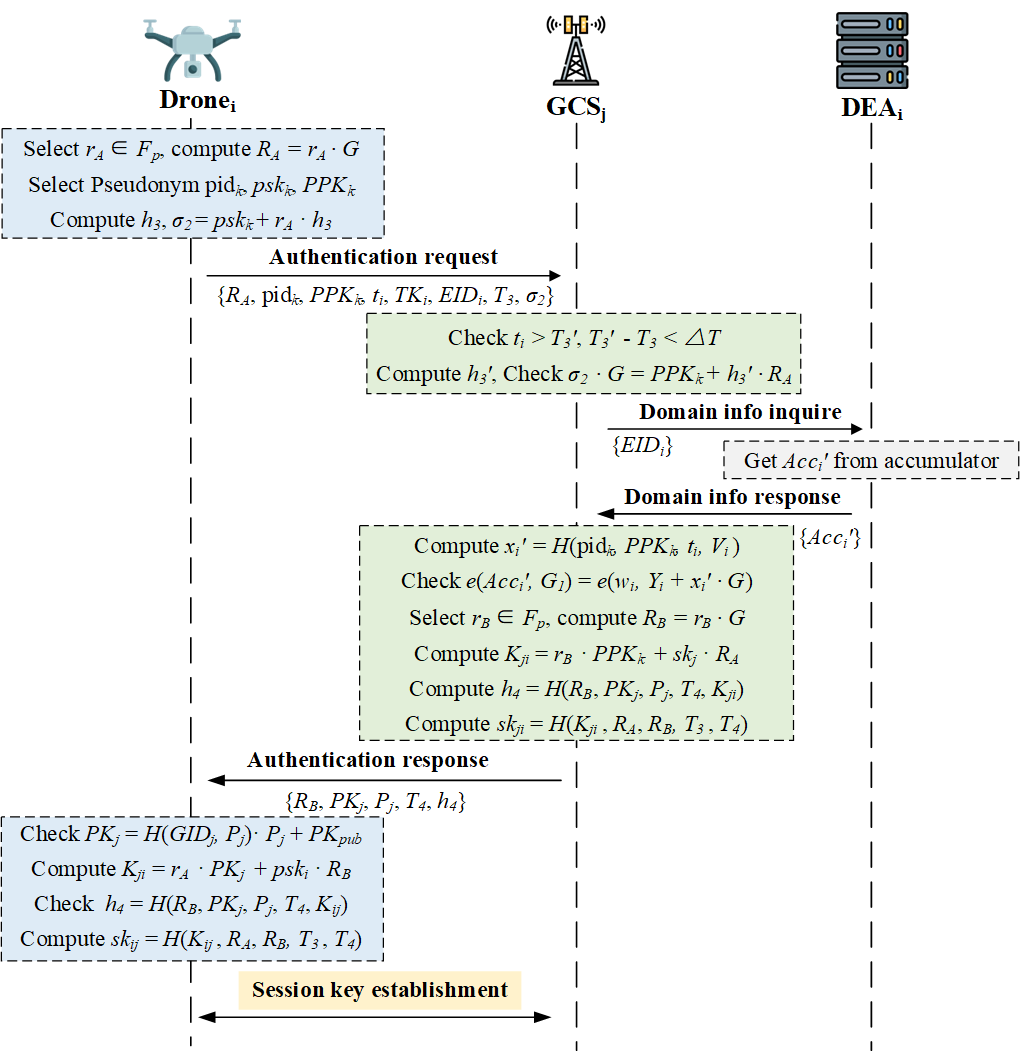}}
\caption{Overview of cross-domain authentication in P\textsuperscript{3}CDA.}
\label{fig}
\end{figure}

\textbf{Step 1:} $\text{D}_{i}$ selects a random number $r_A \in \mathbb{F}_p$ and computes $R_A = r_A \cdot G$. Next, $\text{D}_{i}$ selects an authorized pseudonym at the current moment and computes $h_3 = H(R_A, \text{pid}_k, {PPK}_k, t_i, TK_i, EID_i, T_3, GID_j)$, $\sigma_2 = {psk}_k + r_A \cdot h_3$, where $GID_j$ represents the identity of $\text{GCS}_{j}$. After these computation, $\text{D}_{i}$ sends a cross-domain authentication request to $\text{GCS}_{j}$, including $(R_A, \text{pid}_k, {PPK}_k, t_i, TK_i, EID_i, T_3, \sigma_2)$.

\textbf{Step 2:} Upon receiving the cross-domain authentication request, $\text{GCS}_{j}$ first records the current timestamp $T_3'$, ensuring that $t_i > T_3'$ and $T_3' - T_3 \leq \Delta T$. Next, $\text{GCS}_{j}$ computes $h_3' = H(R_A, \text{pid}_k, {PPK}_k, t_i, TK_i, EID_i, T_3, GID_j)$, and verifies whether $\sigma_2 \cdot G = PPK_k + h_3' \cdot R_A$. If this check passes, $\text{GCS}_{j}$ further verifies the authorization token $TK_i$ by retrieving the accumulator information $Acc_i'$ of domain $i$ using $EID_i$. It then computes $x_i' = H(\text{pid}_k, {PPK}_k, t_i, V_i)$, and verifies whether $e(Acc_i', G_1) = e(w_i, Y_i + x_i' \cdot G_1)$. If the validation is successful, $\text{GCS}_{j}$ confirms that $\text{D}_{i}$ is a legitimate drone. $\text{GCS}_{j}$ then prepares a response message, selecting a random number $r_B \in \mathbb{F}_p$ and computing $
R_B = r_B \cdot G$. Lastly, $\text{GCS}_{j}$ computes the shared-key $K_{ji} = r_B \cdot PPK_k + sk_j \cdot R_A$. After computation, $\text{GCS}_{j}$ sends a response message that includes $(R_B, PK_j, P_j, T_4, h_4)$, where $h_4 = H(R_B, PK_j, P_j, T_4, K_{ji})$.

\textbf{Step 3:} Upon receiving the response message, $\text{D}_{i}$ first checks the correctness of $T_4$ to prevent replay attacks. Next, $\text{D}_{i}$ verifies the correctness of $\text{GCS}_{j}$'s public key by checking whether $PK_j = H(GID_j, P_j) \cdot P_j + PK_{\text{pub}}$. If this verification passes, $\text{D}_{i}$ computes the shared-key $K_{ij} = r_A \cdot PK_j + {psk}_k \cdot R_B$, and verifies whether $h_4 = H(R_B, PK_j, P_j, T_6, K_{ij})$. After these verification, both parties generate the session key $sk_{ij} = H(K_{ij}, R_A, R_B,T_3, T_4)$.

\subsection{Pseudonym Revocation Phase}
To prevent malicious drones from undermining system security, P\textsuperscript{3}CDA incorporates a revocation phase that removes malicious drones and their associated authorization pseudonyms. 

\textbf{Step 1:} If $\text{D}_{i}$ exhibits misbehavior during cross-domain interactions, $\text{GCS}_{j}$ submits a misbehavior report to the TA. This report includes evidence of $\text{D}_{i}$'s malicious behaviors along with the relevant information provided by $\text{D}_{i}$ during the cross-domain interaction.

\textbf{Step 2:} TA verifies the correctness of the evidence provided by $\text{GCS}_{j}$. If it is confirmed that $\text{D}_{i}$ has engaged in malicious behaviors, TA restores its identity using the token parameter, $ID_i' = V_i \oplus H(sk_{\text{pub}} \cdot PK_{\text{ETA}}^i + H(\text{pid}_i, PPK_i) \cdot G)$. TA then removes the registration information of $ID_i'$ from the registration list, and adds $\text{D}_{i}$'s identity information $ID_i$ to the system's revocation list, effectively revoking it from the IoD system. Additionally, TA computes $z_i = H(\text{pid}_i, PPK_i, t_i, V_i)$, and issues a pseudonym revocation request about $\text{D}_{i}$ by sending $(ID_i', z_i, w_i)$ to $\text{DEA}_{i}$.

\textbf{Step 3:} When $\text{DEA}_{i}$ receives the revocation request, it first verifies whether $Acc_i = (y_i + z_i) \cdot w_i$ is satisfied. If the validation passes, $\text{DEA}_{i}$ proceeds to revoke the authorization pseudonym about this drone by updating the accumulator information for domain $i$. Specifically, $\text{DEA}_{i}$ then updates the accumulator value $Acc_i' = \frac{1}{(y_i + z_i)} \cdot Acc_i$, and broadcasts this authorization revocation operation within domain $i$, including $
(Acc_i', z_i)$.

\textbf{Step 4:} After receiving the revocation operation, other drones in the domain $i$ must update the token parameters for cross-domain authorization pseudonyms. For the authorized pseudonym information $(\text{pid}_j, \text{PPK}_j, t_j, V, w_j)$, they compute $x_j = H(\text{pid}_j, \text{PPK}_j, t_j, V)$, and update witness $w_j' = \frac{1}{z_i - x_j} \cdot w_j + \frac{1}{x_j - z_i} \cdot Acc_i'$. The revocation process for the identity and its authorized pseudonym of the malicious drone $\text{D}_{i}$ is successfully completed within the system.

\subsection{Pseudonym Update Phase}
As shown in Fig. 2, we have adopted a hierarchical approach to achieve efficient batch pseudonym registration and update. Layer 2 is used by the drone to locally generate pseudonym information, while Layer 1 is handled by the TA, which processes the drone's registration or update requests and generates public parameters for the drone. These parameters are used to verify the correctness of pseudonym data blocks, including the Merkle root and associated key material.

In the pseudonym update phase, we utilize chameleon hashing to generate the Merkle root node, allowing the drone to efficiently re-generate batches of pseudonyms. When the expiration of $TP_1$ in Layer 1 occurs, all pseudonym blocks in the previously constructed MHT become invalid. At this phase, the drone adjusts the number of pseudonyms required for the next period based on its resource usage in the previous period. Specifically, the drone regenerates the pseudonym data block $m_i' = (\text{pid}_i', \text{PPK}_i', t_i')$ and reconstructs the entire pseudonym MHT. The drone then sends a pseudonym update request to the TA to refresh the Merkle root value, including $(LR_1', LR_2', TP_2)$. The TA uses its private key $sk_{\text{pub}}$ and the adapt function of the chameleon hash to compute a new hash collision for $h_{\text{root}}$. It then updates the authentication parameter to $(r', K')$, where $K' = k' \cdot G$ and $r' = r_{\text{root}} - H(LR_1', LR_2', K') \cdot (k' + sk_{\text{TA}})$. Notably, with the assistance of hash collisions, the drone does not need to re-calculate its key information after a pseudonym update.

When used for cross-domain authentication, to ensure strong unlinkability, the drone must perform pseudonym updates at a small time scale (Layer 2). The drone can update its cross-domain authorization pseudonym by re-executing the intra-domain login phase. Additionally, since authorization pseudonyms and tokens from previous time periods expire and cannot pass validation, the drone can simply discard them, thereby reducing storage overhead.


\section{Security Analysis}
In this section, we demonstrate the security of P\textsuperscript{3}CDA through formal proofs, the ProVerif tool, and heuristic analysis.

\subsection{Formal Security Proof}
We first provide a detailed security proof under the CK adversary model using the sequence-of-games approach. The CK model characterizes the adversary’s capabilities through a set of queries, as detailed below.

\textbf{Players.} We consider a three-party interaction scenario involving drones, GCS, and DEA. Each participant can initiate multiple sessions to represent independent executions of the scheme, which may run concurrently. We denote an instance of a drone, a ground control server, and an edge registration agent as $D^i$, ${GCS}^j$, and ${DEA}^k$, respectively. In addition, each of the above instances is collectively denoted by $I^s$ (i.e., $I \in D \cup GCS \cup DEA$).

\textbf{Queries.} The types of queries available to the adversary $\mathcal{A}$ are defined as follows:

\begin{itemize}

\item{${Execute}(D^i, {GCS}^j, {DEA}^k)$: This query simulates passive attacks, allowing $\mathcal{A}$ to access all messages transmitted during the honest execution of the scheme.}

\item{${Send}(I^s,m)$: This query simulates active attacks, allowing $\mathcal{A}$ to modify, delete, or replay messages. Specifically, $\mathcal{A}$  may send a message $m$ to any instance $I^s$, which will respond if the message is valid and ignore it otherwise.}

\item{${Reveal}(I^s)$: This query simulates session key compromise, allowing $\mathcal{A}$ to obtain the session key accepted and stored by any instance $I^s$.}

\item{${Corrupt}(D^i, {GCS}^j)$: This query simulates a long-term key compromise, enabling $\mathcal{A}$ to access the long-term key stored by instances $D^i$ and ${GCS}^j$. The specific outputs for the two types of instances are as follows.}

\begin{itemize}

\item{When $I^s=D^i$, this query outputs the long-term keys stored in $\text{D}_{i}$, including $\{sk_i,PK_i\}$ and $\{psk_k,PPK_k\}, k \in \{ 1,N \}$.}

\item{When $I^s={GCS}^j$, this query outputs the long-term keys stored in GCS, including $\{sk_j,PK_j\}$.}

\end{itemize}

\item{${Test}(D^i, {GCS}^j)$: This query is used to evaluate the semantic security of the scheme. The adversary $\mathcal{A}$ may launch this query to instance $D^i$ or ${GCS}^j$. If the target instance has not yet established a session key, or if the freshness notion is not met, it returns $\perp$. Otherwise, the instance flips a random bit $b$; if $b=1$, the session key is returned; If $b=0$, a random value of the same length as the session key is returned.}

\end{itemize}

\textbf{Partnering.} In the CK adversary model, partnering is defined based on session identifiers \cite{b11}. A participant initiates a session by executing a protocol instance with parameters $(P_i, P_j, s, \text{role})$, where $\text{role} \in \{\text{initiator}, \text{responder}\}$. Two sessions $(P_i, P_j, s, \text{role})$ and $(P_j, P_i, s', \text{role}')$ are considered matching if they share the same session identifier $s=s'$ and $\text{role}' \ne \text{role}$.

\textbf{Semantic Security.} In authentication and key agreement protocols, session key confidentiality is a fundamental security goal. The adversary $\mathcal{A}$ is allowed to ask a polynomial number of queries during the protocol execution, including $Execute$, $Send$, and $Reveal$ queries. For any instance that satisfies the freshness notion, $\mathcal{A}$ can also initiate a $Test$ query to challenge the confidentiality of the session key. After the $Test$ query, $\mathcal{A}$ attempts to guess the hidden bit $b$. If the guess $b'$ matches the actual bit $b$, the adversary is considered to have won the game. This event is denoted as $\mathsf{Succ}_A$. Formally, the adversary’s advantage in breaking the semantic security of the scheme is defined as:
\begin{equation}
\mathsf{Adv}^{\mathsf{ake}}_{\mathcal{P}} = 2 \Pr[\mathsf{Succ}_\mathcal{A}] - 1.
\end{equation}

\textbf{Freshness.} Since $\mathcal{A}$ could trivially win the security game by obtaining session keys through $Reveal$ queries, the freshness notion is introduced to restrict such access. An instance $I^s$ is considered fresh if it satisfies the following conditions: 1) $I^s$ has successfully accepted a session key; 2) Neither $I^s$ nor its partner instance has been asked to a $Reveal$ query; 3) $Corrupt$ query is asked at most once.

\textbf{Theorem 1.} \textit{Let the P\textsuperscript{3}CDA scheme $\mathcal{P}$ run over a group of prime order $p$, with the hash function producing $l$-bit outputs. Let $\mathcal{A}$ be an adversary attempting to break the semantic security of the scheme within a time bound $t$. During its interaction with the instances, $\mathcal{A}$ performs $q_s$ $Send$ queries, $q_e$ $Execute$ queries, and $q_h$ $Hash$ queries. Then we obtain:}
\begin{IEEEeqnarray}{rCl}
\mathit{Adv}^{\mathsf{ake}}_{\mathcal{P}} & \leq & 2 (q_s+q_e)\mathit{Adv}^{\mathit{q-SDH}}(t) + 6 q_h \cdot \mathit{Adv}_{\mathcal{P}}^{\mathit{ECCDH}}(t') \nonumber\\
&& + \frac{q_h^2 + 2 q_{s}}{2^l} + \frac{3(q_{s} + q_{e})^2 + 17q_{s}}{3p}.
\end{IEEEeqnarray}

\textit{Where $t' \leq t + (q_{s} + q_{e} + 1) \cdot T_m
$, $T_m$ denotes the time required for scalar multiplication operations on $\mathbb{G}$.}

\begin{IEEEproof}
To prove Theorem 1, we adopt the sequence-of-games approach \cite{b13}, constructing a sequence of security games with incremental modifications. In each successive game, the adversary's advantage is reduced step by step, eventually approaching zero in the final game $G_8$. In addition, we define the following events for each game.

\begin{itemize}

\item{\textbf{$\mathsf{Succ}_n$:} The advantage of adversary $\mathcal{A}$ successfully guessing the bit $b$ in $Test$ query.}

\item{\textbf{$\mathsf{AskAuth}_n$:} The adversary $\mathcal{A}$ correctly computes $K_{ij}$ and asks a $Hash$ query on $(R_B, PK_j, P_j, T_4, K_{ij})$.}

\item{\textbf{$\mathsf{AskHash}_n$:} The adversary $\mathcal{A}$ asks a $Hash$ query about $(K, T_3, T_4, R_A, R_B)$, where $K$ is $K_{ij}$ or $K_{ji}$.}

\end{itemize}

\textbf{Game $G_0$:} This game models the real attack by adversary $\mathcal{A}$ in our scheme. According to the definition of semantic security, the adversary's advantage is given by
\begin{equation}
\mathsf{Adv}^{\mathsf{ake}}_{\mathcal{P}} = 2 \Pr[\mathsf{Succ}_{0}] - 1.
\end{equation}

\textbf{Game $G_1$:} In this game, we simulate a hash oracle (modeled as a random oracle) $\mathcal{H}$ that responds to $Hash$ queries by maintaining two lists: $\mathcal{L}_\mathcal{H}$ and $\mathcal{L}_\mathcal{A}$. Specifically, $\mathcal{L}_\mathcal{H}$ records all $Hash$ queries and their corresponding responses, while $\mathcal{L}_\mathcal{A}$ specifically logs the queries made by the adversary $\mathcal{A}$. Additionally, we construct a private random oracle, denoted as $\mathcal{H}'$, which will be used in Game $G_7$. It can be found that there is no difference in this game compared to $G_0$, and thus the adversary's advantage does not increase.
\begin{equation}
\left| \Pr[\mathsf{Succ}_1] - \Pr[\mathsf{Succ}_0] \right| = 0.
\end{equation}

\textbf{Game $G_2$:} We eliminate potential collisions between messages from different sessions, including both random number collisions and $Hash$ query collisions. If any collision is detected, the game is immediately aborted. According to the birthday paradox, the probability of a collision is bounded by $ \frac{q^2}{2} \cdot 2^{-l}$, where $q$ denotes the number of queries. Specifically, the probability of random number collisions between messages exchanged by different instances is bounded by $ \frac{(q_s + q_e)^2}{2p}$. For $Hash$ query collisions, the probability is bounded by $ \frac{q_h^2}{2^{l+1}}$. Thus, we obtain: 
\begin{equation}
\left| \Pr[\mathsf{Succ}_2] - \Pr[\mathsf{Succ}_1] \right| \leq \frac{(q_s + q_e)^2}{2p} + \frac{q_h^2}{2^{l+1}}.
\end{equation}

\textbf{Game $G_3$:} In this game, we exclude the possibility that the adversary $\mathcal{A}$ successfully guesses the verification parameters without querying the hash oracle, including $({TK_i}^*,{\sigma_2}^*)$ in $(R_A, \text{pid}_k, {PPK}_k, t_i, {TK_i}^*, EID_i, T_3, {\sigma_2}^*)$ and ${h_4}^*$ in $(R_B, PK_j, P_j, T_4, {h_4}^*)$. Given that message collisions have been eliminated in the previous game, the probability of correctly guessing ${w_i}^*,{\sigma_2}^*$  in the authentication request is bounded by $\frac{q_s}{3p}$, and guessing ${h_4}^*$ in the authentication response is bounded by $\frac{q_s}{2^{l}}$. Accordingly, we modify the instances' interaction process as follows:

\begin{itemize}

\item{After receiving the authentication request message, ${GCS}^j$ first checks whether the tuples $\{0,(R_A, \text{pid}_k, {PPK}_k, t_i, {TK_i}^*, EID_i, T_3,GID_j),*\}$ exists in $\mathcal{L}_\mathcal{A}$, and $\{0,(R_A, \text{pid}_k, {PPK}_k, t_i),*\}$ exists in $\mathcal{L}_\mathcal{H}$. If no matching records are found, the game is terminated; otherwise, ${GCS}^j$ proceeds with the authentication process.}

\item{After receiving the authentication reponse message, $D^i$ checks whether the tuples $\{0,(*),{h_4}^*\}$ exists in $\mathcal{L}_\mathcal{A}$. If no matching record is found, the game is terminated; otherwise, $D^i$ proceeds with the authentication process.}

\end{itemize}

Games $G_3$ and $G_2$ are indistinguishable unless the adversary $\mathcal{A}$ correctly guesses the verification parameters and constructs a valid authentication message. Thus, we obtain:
\begin{equation}
\left| \Pr[\mathsf{Succ}_3] - \Pr[\mathsf{Succ}_2] \right| \leq \frac{q_s}{3p} + \frac{q_s}{2^{l}}.
\end{equation}

\textbf{Game $G_4$:} In this game, we exclude the possibility of the adversary $\mathcal{A}$ successfully guessing the shared key $K_{ij}$, which could be achieved through the following three approaches.

\begin{itemize}

\item{$\mathcal{A}$ may attempt a direct guess of the shared key, which succeeds with probability at most $\frac{q_s}{2p}$.}

\item{$\mathcal{A}$ may obtain the drone’s long-term secret ${psk}_k$ via sending a ${Corrupt}(D^i)$ query and then attempt to guess the shared-key by guessing $r_A$, which succeeds with probability at most $\frac{q_s}{p}$.}

\item{$\mathcal{A}$ may obtain the drone’s long-term secret ${sk}_j$ via sending a ${Corrupt}({GCS}^j)$ query and then attempt to guess the shared-key by guessing $r_B$, which succeeds with probability at most $\frac{q_s}{p}$.}

\end{itemize}

Games $G_4$ and $G_3$ are indistinguishable unless $\mathcal{A}$ successfully guesses the shared key $K_{ij}$, which we can get:
\begin{equation}
\left| \Pr[\mathsf{Succ}_4] - \Pr[\mathsf{Succ}_3] \right| \leq \frac{5q_s}{2p}.
\end{equation}

\textbf{Game $G_5$:} In this game, we exclude the possibility that adversary $\mathcal{A}$ can forge a valid authorization token $TK_i$. If $\mathcal{A}$ is able to successfully forge such a token, then a reduction algorithm $\mathcal{R}$ can be constructed to solve the q-SDH assumption by leveraging $\mathcal{A}$'s capabilities. The details of this reduction can be found in the original paper on the accumulator construction \cite{b14}. Therefore, we obtain:
\begin{equation}
\left| \Pr[\mathsf{Succ}_5] - \Pr[\mathsf{Succ}_4] \right| \leq (q_s+q_e)\mathit{Adv}^{\mathit{q-SDH}}(t).
\end{equation}

\textbf{Game $G_6$:} In this game, we eliminate the possibility that the adversary correctly computes the verification parameter $h_4$ by aborting the game. To achieve this, we modify the instance's simulation.

\begin{itemize}

\item{When $D^i$ receives an authentication response message, it will perform the defined validation process. If the message passes the verification, $D^i$ will check whether $\{0,(R_B, PK_j, P_j, T_4, K_{ij}),h_4\}$ exists in the $\mathcal{L}_\mathcal{A}$. If the record exists, the game is aborted; otherwise $D^i$ computes the session key $sk_{ij}$.}

\end{itemize}

Games $G_6$ and $G_5$ are indistinguishable unless the event $\mathsf{AskAuth}_6$ occurs, which we can get:
\begin{equation}
\left| \Pr[\mathsf{Succ}_6] - \Pr[\mathsf{Succ}_5] \right| \leq \Pr[\mathsf{AskAuth}_6].
\end{equation}

\textbf{Game $G_7$:} In this game, we replace the hash oracle $\mathcal{H}$ with the private hash oracle $\mathcal{H}'$ to compute some critical parameters.
\begin{align}
h_4 &= \mathcal{H}'(R_B, PK_j, P_j, T_4) ,\\
sk_{ji} &= sk_{ji} = \mathcal{H}'(R_A, R_B, T_3, T_4).
\end{align}

Since the adversary $\mathcal{A}$ has no access to the oracle $\mathcal{H}'$, Games $G_7$ and $G_6$ are indistinguishable unless the event $\mathsf{AskHash}_7$ occurs, which we can get:

\begin{align}
\left| \Pr[\mathsf{Succ}_7] - \Pr[\mathsf{Succ}_6] \right| \leq \Pr[\mathsf{AskHash}_7], \\
\left| \Pr[\mathsf{AskAuth}_7] - \Pr[\mathsf{AskAuth}_6] \right| \leq \Pr[\mathsf{AskHash}_7].
\end{align}

At this point, all feasible strategies for adversary $\mathcal{A}$ to successfully guess $b$ have been excluded, leaving random guessing as the only remaining option. Moreover, while $\mathcal{A}$ may obtain other session keys via ${Reveal}(D^i)$ or ${Reveal}({GCS}^j)$ queries, these do not improve its successful probability, as the session keys are generated using a private hash oracle and are independent of the $K_{ij}$. So we can get:
\begin{equation}
\Pr[\mathsf{Succ}_7] = \frac{1}{2}.
\end{equation}

\textbf{Game $G_8$:} In this game, we leverage the random self-reducibility of the ECCDH assumption and adapt the instance's simulation accordingly. Given an ECCDH instance $(R_A,PK_j)$, the instance does not need access to $r_A$ and $sk_j$ to correctly simulate the authentication process.  Therefore, Games $G_7$ and $G_8$ are computationally indistinguishable, i.e.:
\begin{equation}
\Pr[\mathsf{AskH}_8] = \Pr[\mathsf{AskH}_7].
\end{equation}

If event $\mathsf{AskH}_8$ occurs, it indicates that the adversary $\mathcal{A}$ has correctly computed $K_{ij}$ or $K_{ji}$ and issued $Hash$ queries containing these values. In such case, we can extract from the list $\mathcal{L}_\mathcal{A}$ as a solution to the ECCDH assumption $(R_A,PK_j,CDH(R_A,PK_j))$ with probability at least $\frac{1}{q_h}$, which $CDH(R_A,PK_j)=K_{ij}-{psk}_k \cdot R_B=K_{ji}-r_B \cdot PPK_k$. Therefore, we obtain: 
\begin{equation}
\left| \Pr[\mathsf{AskH}_8] \right| \leq q_h \cdot \mathit{Adv}_{\mathcal{P}}^{\mathit{ECCDH}}(t').
\end{equation}

\textbf{Conclusion of the proof:} At the end, by combining Eqs.(3)-(16), we obtain the overall advantage of adversary $\mathcal{A}$ as:

\begingroup
\setlength{\abovedisplayskip}{4pt}
\setlength{\belowdisplayskip}{4pt}
\begin{IEEEeqnarray}{rCl}
\mathit{Adv}^{\mathsf{ake}}_{\mathcal{P}} & = & 2\Pr[\mathsf{Succ}_7] - 1 + 2\left(\Pr[\mathsf{Succ}_0] - \Pr[\mathsf{Succ}_7]\right) \notag\\
& \leq & 2(q_s + q_e)\mathit{Adv}^{\mathit{q\text{-}SDH}}(t) 
+ 6q_h\cdot\mathit{Adv}_{\mathcal{P}}^{\mathit{ECCDH}}(t') \notag\\
&& + \frac{q_h^2 + 2q_s}{2^l}
+ \frac{3(q_s + q_e)^2 + 17q_s}{3p}. \notag
\end{IEEEeqnarray}
\endgroup
\end{IEEEproof}

\subsection{Formal Security Verification}
To further evaluate the security of the proposed scheme, we utilize the ProVerif \cite{b16} tool, which is widely used for the formal analysis of authentication protocols \cite{b17,b18,b19}. ProVerif analyzes protocols modeled in the applied pi-calculus and can verify security properties for an unbounded number of messages and sessions. ProVerif is based on the Dolev-Yao adversary model and supports formal verification of security properties such as confidentiality, reachability, and authentication. We implement the proposed scheme in the ProVerif tool, and the code is available in GitHub \footnote{\url{https://github.com/chengqi1223/P3CDA}}       .

To accurately model the proposed scheme in ProVerif, we define two communication channels: a private channel and a public channel. The private channel is assumed to be secure and inaccessible to the adversary, and is employed solely during the identity registration phase, while all other message exchanges occur over the public channel. We define three data types: string, point, and API (representing Merkle path proofs), and model the cryptographic primitives used in the scheme, including hash functions, XOR, elliptic curve point addition, scalar multiplication, etc., using rewrite rules. Notably, the ECDH protocol is implemented via the equation rule.

Based on the above definitions, we simulated the execution process of each entity in the scheme, including the drone, GCS, DEA, and TA. Given that all DEAs are functionally equivalent and capable of secure mutual communication, we modeled a single DEA to simplify the simulation. This does not affect the analysis results. We defined three security queries to verify confidentiality and anonymity properties: the secrecy of the session keys $sk_{ij}$ and $sk_{ji}$, and the anonymity of the identity $ID_i$. Additionally, to verify authentication properties, we defined 10 events and 5 queries to ensure that entities can achieve secure mutual authentication across the identity registration, intra-domain login, and cross-domain authentication phases.

Based on the input protocol, the ProVerif performs symbolic analysis by simulating attacks to determine whether the scheme meets the defined security goals. If a goal is not met, ProVerif returns `\textit{false}' and attempts to reconstruct and output the corresponding attack trace; otherwise, it returns `\textit{true}'. As shown in Fig. 6, the simulation results confirm that the proposed scheme achieves secure mutual authentication among entities while ensuring identity anonymity and the confidentiality of the session keys.

\begin{figure}
\centerline{\includegraphics[width=0.49\textwidth]{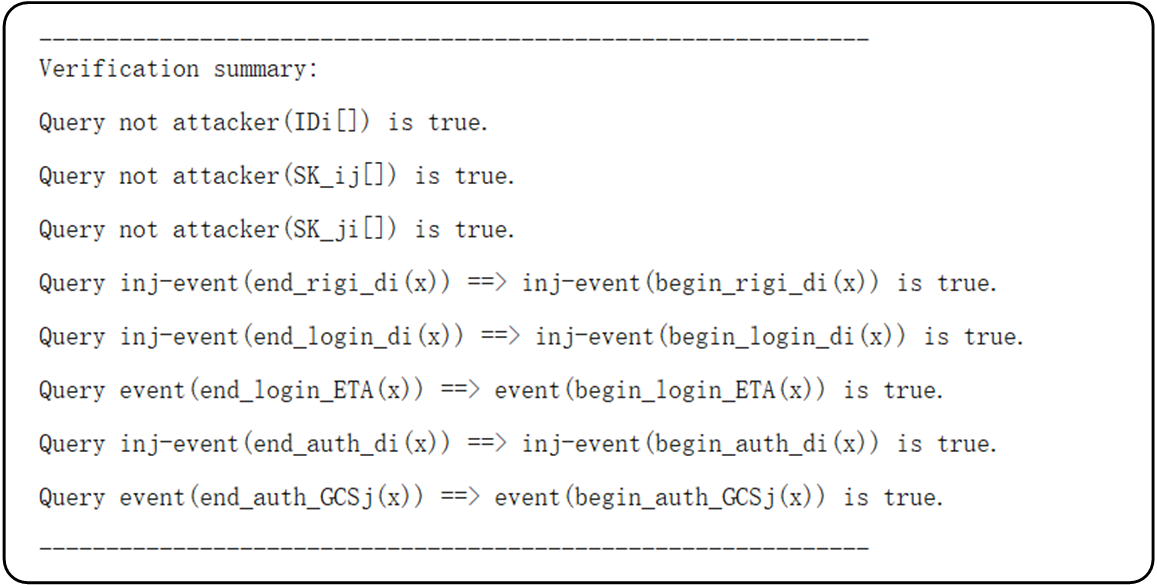}}
\caption{Security verification results using the ProVerif tool.}
\label{fig}
\end{figure}

\subsection{Heuristic Analysis}
This section presents a heuristic analysis demonstrating how P\textsuperscript{3}CDA meets the defined security goals.

\subsubsection{Mutual Authentication}
In $\text{D}_{i}$'s intra-domain login request, the DEA verifies its legitimacy using IBC by checking the correlation between the public key and the identity ${ID}_i$, Merkle proof $\pi$, and Merkle root $h_{\text{root}}$. Meanwhile, the binding between the pseudonym and the identity is verified through the parameter $\sigma_1$. $\text{D}_{i}$ then verifies the DEA’s reply message by checking the hash value $h_2 = H(TK_i, T_2, r_s')$, where $r_s'$ is the shared secret between $\text{D}_{i}$ and $\text{DEA}_{i}$. In the cross-domain authentication request, $\text{GCS}_{j}$ can verify the legitimacy of $\text{D}_i$'s identity by checking the membership witness $w_i$ in the accumulator $Acc_i$ for the pseudonym information $x_i = H(\text{pid}_i, PPK_i, t_i, V_i)$, as it has already been authorized.

\subsubsection{Secure Session Key}
During the cross-domain authentication process, $\text{D}_i$ and $\text{GCS}_{j}$ independently compute the same shared-key $K_{ij} = r_A \cdot PK_j + {psk}_k \cdot R_B = r_B \cdot PPK_k + sk_j \cdot R_A = K_{ji}$, from which they derive the session key $sk_{ij}$ and $sk_{ji}$. Since the random values $r_A$, $r_B$, and the secret keys ${psk}_k$, ${sk}_j$ are inaccessible to the attacker, the scheme can securely and reliably generate a session key between the authenticated parties.

\subsubsection{Conditional Anonymity}
In P\textsuperscript{3}CDA, $\text{D}_i$ is not required to transmit any information related to its identity over the public channel. In the intra-domain login phase, $\text{D}_i$ hides its identity by computing $({ID}_i || P_i) \oplus H(\text{pid}_k || {psk}_k \cdot PK_{\text{ETA}}^i)$. Since ${psk}_k \cdot PK_{\text{ETA}}^i$ is the shared-secret between $\text{D}_i$ and $\text{DEA}_{i}$, only $\text{DEA}_{i}$ can correctly recover $\text{D}_i$'s identity from it. During the cross-domain authentication phase, $\text{D}_i$ uses authorized pseudonyms for authentication, preventing the attacker from obtaining its identity in this process. Meanwhile, to prevent malicious drones from compromising system reliability, a critical parameter $V_i$ is added in $\text{D}_i$'s pseudonym authorization token. If necessary, the TA can recover the identity of the malicious drone by computing $V_i \oplus H(sk_{\text{pub}} \cdot PK_{\text{ETA}}^i + H(\text{pid}_i, PPK_i) \cdot G)$. In addition, since the authorization information added to the accumulator is bound in the form of $x_i = H(\text{pid}_i, PPK_i, t_i, V_i)$, malicious drones cannot tamper with $V_i$ to evade identity traceability.

\subsubsection{Unlinkability}
We employ the Merkle hash tree to verify the integrity of pseudonym information. However, the Merkle path proofs of multiple intra-domain login requests may include repeated nodes. To prevent attackers from linking pseudonyms through these repeated path nodes, the shared secret $r_s'$ between $\text{D}_i$ and $\text{DEA}_{i}$ is used to hide the path nodes, ensuring that only $\text{DEA}_{i}$ can correctly reconstruct the path proof. Additionally, to enhance unlinkability, the scheme supports a batch pseudonym update mechanism that ensures drones use distinct pseudonyms for each cross-domain authentication, effectively preventing identity linkage.

\subsubsection{Forward Secrecy}
Consider an attacker who has obtained all the long-term secrets stored by the $\text{D}_{i}$ or $\text{GCS}_{j}$, including $\{(sk_i,psk_1, \dots, {psk}_N), sk_j\}$, as well as the parameters $R_A$ and $R_B$ by eavesdropping on past communication sessions. In our scheme, the session key ${sk}_{ij}$ is computed by $H(K_{ij}, R_A, R_B,T_3, T_4)$. The attacker must correctly derive the shared key $K_{ij}$ or $K_{ji}$, which depends on the random numbers $r_A$ and $r_B$. These values are generated locally by $\text{D}_{i}$ and $\text{GCS}_{j}$, respectively, and are erased after the session, rendering them inaccessible to the attacker. Therefore, our scheme ensures strong forward secrecy.

\subsubsection{Resisting Man-in-the-Middle Attack}
For an attacker attempting to impersonate a legitimate entity, the use of IBC and cryptographic accumulators makes it infeasible to forge another entity's identity to pass verification. Meanwhile, since all messages in the scheme include integrity verification parameters (e.g., $\sigma_1$, $h_2$, $\sigma_2$, $h_4$, etc.), an attacker cannot tamper with any message in a way that would make it acceptable to the verifier.

\subsubsection{Resisting Replay Attack}
In our scheme, all transmitted messages include timestamps (e.g., $T_1$, $T_2$, etc.), allowing the receiver to verify message freshness by checking whether the time interval from the current moment is within the allowed threshold $\Delta T$. Furthermore, since attackers cannot tamper with message contents, they cannot launch replay attacks by forging or altering timestamps.

\subsubsection{Resisting Privileged Insider Attack}
We also consider the attacker with internal privileges who has access to the system parameters $\{G, G_1, PK_{\text{pub}}, EID_i, PK_{\text{ETA}}^i, Y_i, Acc_i\}$, and even to sensitive information stored within the TA except for the private key, such as $\{r_{DA}, P_i, w\}$. Even so, the attacker is still unable to derive $\text{D}_{i}$'s private key, thereby compromising the security of the scheme. This is because the computation of $sk_i$ relies on the secret $r_i$, which is held by $\text{D}_{i}$, and the pseudonym key ${psk}_i$ is also generated entirely on the drone. Both values are inaccessible to the attacker. Therefore, the scheme can be resilient against privileged insider attacks.

\section{Performance Evaluation}
This section describes the construction of the experimental environment, followed by theoretical analysis and simulation evaluation to demonstrate the effectiveness of P\textsuperscript{3}CDA.

\begin{table}[t]
\centering
\caption{Time Cost for Cryptography Operations (Unit: ms)}
\label{tab:time_cost}
\renewcommand{\arraystretch}{1.2}
\begin{tabular}{c|c c c c c c}
\specialrule{1.1pt}{1.1pt}{1.1pt} 
Operations &
$T_{p}$ &
$T_{e}$ &
$T_{m}$ &
$T_{h}$ &
$T_{s}$ &
$T_{v}$ \\
\specialrule{1.1pt}{1.1pt}{1.1pt} 
ETA / GCS &
2.33 & 1.27 & 0.57 & 0.001 & 0.85 & 0.76 \\
Drone &
21.64 & 12.15 & 4.01 & 0.003 & 4.87 & 5.44 \\
\specialrule{1.1pt}{1.1pt}{0pt} 
\end{tabular}
\end{table}

\subsection{Experimental Setup}
To simulate an IoD environment, we used Raspberry Pi 3B devices as drone nodes, each equipped with 1 GB of RAM and a 1.2 GHz ARM Cortex-A53 CPU. Given that the GCS and ETA are ground-deployed with greater computational capabilities, we simulated them using a virtual machine with 8 GB of RAM, a 2.3 GHz Intel Core i7-12700H CPU, and the Ubuntu 22.04 operating system. Based on this setup, we implemented P\textsuperscript{3}CDA in the C language with the Miracl-core library \footnote{\url{https://github.com/miracl/core}}. The security level is set to 128 bits, utilizing the BN254 elliptic curve defined by the equation $Y^2=X^3+3$ over $\mathbb{F}_p$, which supports bilinear pairing operations. Additionally, SHA-256 is used as the hash function.

Table II presents the core cryptographic operations utilized in P\textsuperscript{3}CDA and the comparison schemes. Specifically, ${T}_{{p}}$, ${T}_{e}$, ${T}_{m}$, ${T}_{h}$, ${T}_{s}$, and ${T}_{v}$ represent the execution time for a single bilinear pairing, exponentiation in group $\mathbb{G_T}$, point multiplication in group $\mathbb{G}$, hashing, signature generation, and signature verification, respectively. The ECDSA algorithm is used for digital signatures. It can be observed that, due to their greater computational complexity, the two operations on group $\mathbb{G_T}$ incur significantly higher running times than the other operations. In addition, to more precisely quantify the communication overhead, we define $|\mathbb{G}|$, $|p|$, $|ID|$, and $|T|$ as the lengths of group $\mathbb{G}$, finite field $\mathbb{F}_p$, device identity, and timestamp, respectively. According to the experimental security level settings, these lengths are set to 512, 256, 64, and 64 bits, respectively.

\begin{table*}[!t]
\centering
\caption{Complexity Comparison for the Intra-Domain Login Process}
\label{tab:idl_complexity}
\renewcommand{\arraystretch}{1.2}
\setlength{\tabcolsep}{6.5pt}

\makebox[\textwidth][c]{%
\resizebox{0.95\textwidth}{!}{%
\begin{tabular}{c|cc|cc}
\specialrule{1.2pt}{1.2pt}{1.2pt}
\multirow{2}{*}{Schemes} &
\multicolumn{2}{c|}{Computational Overhead} &
\multicolumn{2}{c}{Communication Overhead} \\
\cmidrule(lr){2-3} \cmidrule(lr){4-5}
& Drone & ETA & Total cost & Rounds \\
\specialrule{1.2pt}{1.2pt}{1.2pt}

CCAP \cite{b3} &
$T_{s}$ &
$2T_{p}+T_{s}+3T_{v}+3T_{m}+2T_{h}$ &
$9|\mathbb{G}|+(4+s)|ID|$ & 4 \\

BAR-CDA \cite{b8} &
$3T_{m}+T_{s}+T_{v}$ &
$13T_{m}+T_{s}+T_{v}+4T_{h}$ &
$9|\mathbb{G}|+4|p|+6|ID|+5|T|$ & 6 \\

BASA \cite{b1} &
$T_{p}+T_{e}+T_{m}+T_{h}$ &
$2T_{p}+2T_{e}+2T_{m}+3T_{h}$ &
$3|\mathbb{G}|+|p|+2|ID|$ & 4 \\

\rowcolor[gray]{0.9}
\textbf{P\textsuperscript{3}CDA} &
$T_{m}+(\log_{2} n+4)T_{h}$ &
$9T_{m}+(\log_{2} n+10)T_{h}$ &
$3|\mathbb{G}|+(\log_{2} n+2)|p|+3|ID|+4|T|$ & 2 \\
\specialrule{1.2pt}{1.2pt}{0pt}
\end{tabular}%
}} 

\vspace{0.5em}
\parbox{0.95\textwidth}{
\footnotesize
\textbullet\ $n$ denotes the number of pseudonyms for the drone; $s$ denotes the number of identities in the revocation list.\\
}
\end{table*}

\begin{table*}[!t]
\centering
\caption{Complexity Comparison for the Cross-Domain Authentication Process}
\label{tab:cda_complexity}
\renewcommand{\arraystretch}{1.2}
\setlength{\tabcolsep}{6.5pt}

\makebox[\textwidth][c]{%
\resizebox{0.95\textwidth}{!}{%
\begin{tabular}{c|ccc|cc}
\specialrule{1.2pt}{1.2pt}{1.2pt} 
\multirow{2}{*}{Schemes} &
\multicolumn{3}{c|}{Computational Overhead} &
\multicolumn{2}{c}{Communication Overhead} \\
\cmidrule(lr){2-4} \cmidrule(lr){5-6}
& Drone & GCS & ETA & Total cost & Rounds \\
\specialrule{1.2pt}{1.2pt}{1.2pt} 

CCAP \cite{b3} &
$2T_{s}+T_{m}+T_{v}$ &
$T_{s}+T_{v}$ &
$17T_{p}+9T_{e}+39T_{m}$ &
$32|\mathbb{G}|+19|p|+7|ID|$ & 12 \\

ER-CDAA \cite{b7} &
$5T_{p}+5T_{e}+15T_{m}$ &
$5T_{p}+5T_{e}+15T_{m}$ &
-- &
$10|\mathbb{G}|+6|p|$ & 4 \\

BAR-CDA \cite{b8} &
$3T_{p}+17T_{m}$ &
$3T_{p}+17T_{m}$ &
-- &
$10|\mathbb{G}|+6|p|$ & 2 \\

BASA \cite{b1} &
$2T_{m}+T_{h}$ &
$2T_{m}+T_{h}$ &
$6T_{p}+6T_{e}+4T_{m}$ &
$10|\mathbb{G}|+12|p|+8|ID|$ & 12 \\

EPPB \cite{b39} &
$8T_{m}+10T_{h}$ &
$8T_{m}+10T_{h}$ &
$4T_{p}+14T_{m}+12T_{h}$ &
$18|\mathbb{G}|+6|p|+2|ID|+6|T|$ & 8 \\

BCAE \cite{b38} &
$6T_{m}+2T_{h}+2T_{v}$ &
$6T_{m}+2T_{h}+2T_{v}$ &
-- &
$10|\mathbb{G}|+|p|+4|ID|$ & 6 \\

\rowcolor[gray]{0.9}
\textbf{P\textsuperscript{3}CDA} &
$4T_{m}+4T_{h}$ &
$2T_{p}+6T_{m}+4T_{h}$ &
-- &
$7|\mathbb{G}|+2|p|+4|ID|+3|T|$ & 4 \\
\specialrule{1.2pt}{1.2pt}{0pt} 
\end{tabular}%
}} 

\end{table*}

\subsection{Theoretical Analysis}

First, we provide a detailed analysis of SOTA cross-domain authentication schemes through theoretical evaluation of their computational and communication overheads. Due to differences in structural design, some schemes do not require intra-domain login. The combined overhead is comprehensively evaluated in the next subsection.

Table III presents a comparison of the complexity of the intra-domain login process.  P\textsuperscript{3}CDA demonstrates the lowest computational overhead on both the drone and ETA sides, with costs of $T_{m}+(\log_{2} n+4)T_{h}$ and $9T_{m}+(\log_{2} n+10)T_{h}$, respectively. BASA incurs significantly higher computational overhead because the drone must perform bilinear pairing-based Identity-based Signature (IBS) operations. Regarding communication overhead, P\textsuperscript{3}CDA requires only two communication rounds and incurs a cost of $3|\mathbb{G}|+(\log_{2} n+3)|p|+3|ID|+4|T|$, ranking second only to BASA. However, BASA involves more communication rounds and relies on the blockchain’s synchronization and consensus processes. Although CCAP and BAR-CDA exhibit comparable computational overhead to P\textsuperscript{3}CDA, they incur significantly higher communication size and require more communication rounds, which limits their suitability for resource-constrained environments.

Table IV summarizes the complexity comparison of the cross-domain authentication process, showing the resource overhead of a single mutual authentication between entities. In addition to the drone and GCS, some schemes also require assistance from the ETA during the authentication process. In terms of computational overhead, P\textsuperscript{3}CDA incurs a lower cost on the drone side at $4T_{m}+4T_{h}$, ranking second only to BASA's $2T_{m}+T_{h}$ and remaining comparable to CCAP's $2T_{s}+T_{m}+T_{v}$. This is particularly important as drones are resource-constrained devices with limited computational capabilities, making it essential to minimize complex cryptographic operations. Furthermore, although BASA and CCAP appear to impose lower computational overhead on the drone side, this is primarily because they offload many intensive bilinear pairing operations to the ETA. However, in large-scale device connectivity scenarios, the ETA must centrally handle high-intensity cryptographic operations, which can lead to significant latency and performance bottlenecks across the IoD system. Moreover, since EPPB, BAR-CDA, and ER-CDAA all require drones to construct NIZKs, these schemes impose significant computational overhead on the drone side. On the GCS side, due to the use of a cryptographic accumulator in P\textsuperscript{3}CDA, the GCS needs to perform two bilinear pairing operations to verify membership proofs. This results in a computational overhead of $2T_{p}+6T_{m}+4T_{h}$, placing it at a moderate level among all the compared schemes. This overhead is acceptable, as GCSs are ground-based with higher computational capacity, and the presence of multiple GCSs within an IoD domain enables task sharing and load balancing.

Furthermore, in terms of communication overhead, P\textsuperscript{3}CDA incurs the lowest total transmission cost, requiring only $7|\mathbb{G}|+2|p|+4|ID|+3|T|$. In contrast, CCAP, EPPB, BAR-CDA, and ER-CDAA require the transmission of anonymous credentials or group signatures along with their corresponding proofs, leading to significantly higher parameter transmission overhead. In addition, BASA also incurs a high total communication size and a large number of communication rounds due to frequent computation offloading. BCAE relies on identity certificates for authentication, which results in slightly higher communication overhead than P\textsuperscript{3}CDA.

\begin{figure}[t]
  \centering
  \begin{subfigure}[b]{0.49\textwidth}
    \includegraphics[width=\textwidth]{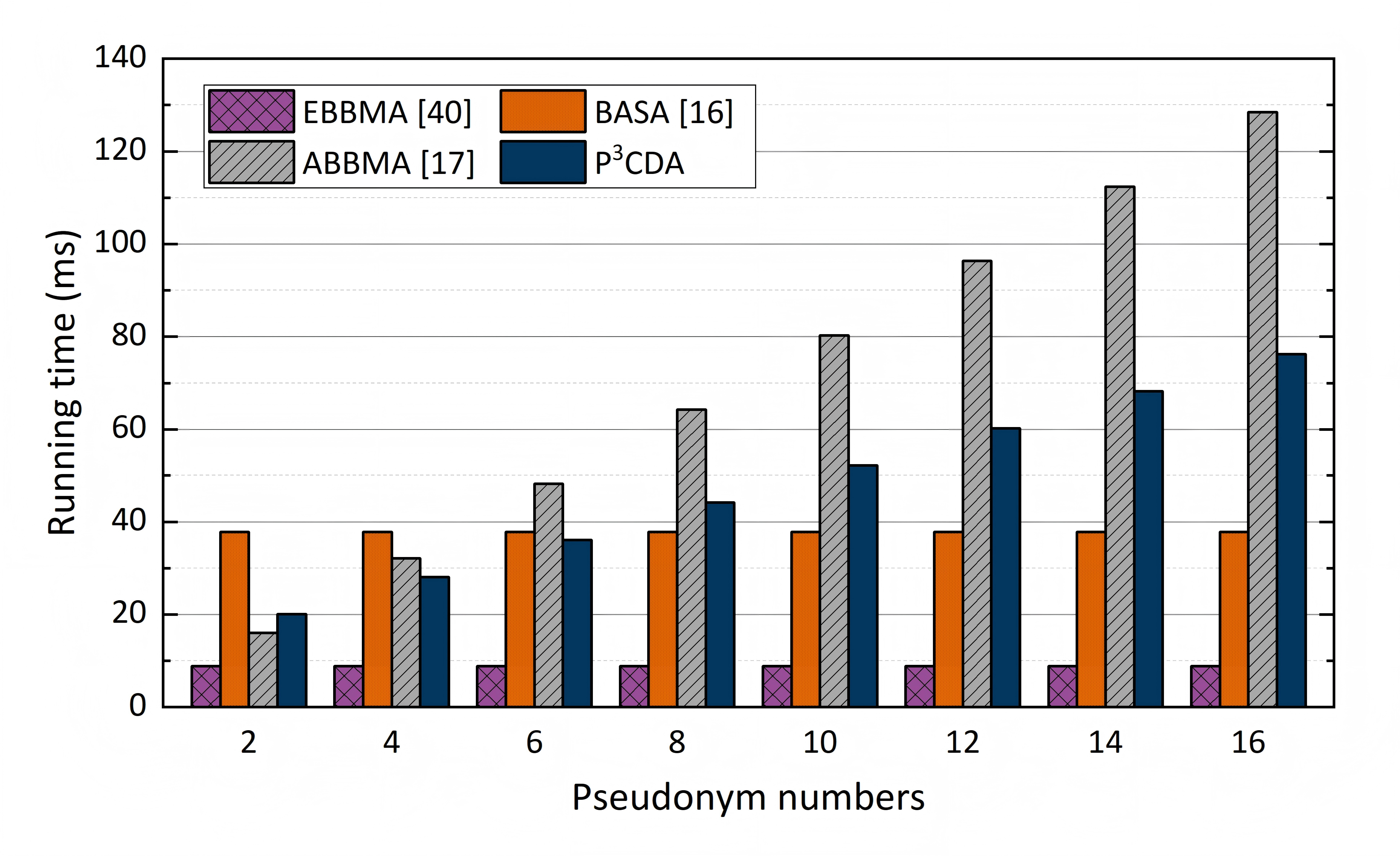}
    \caption{The running time on the drone side}
  \end{subfigure}
  \hfill
  \begin{subfigure}[b]{0.49\textwidth}
    \includegraphics[width=\textwidth]{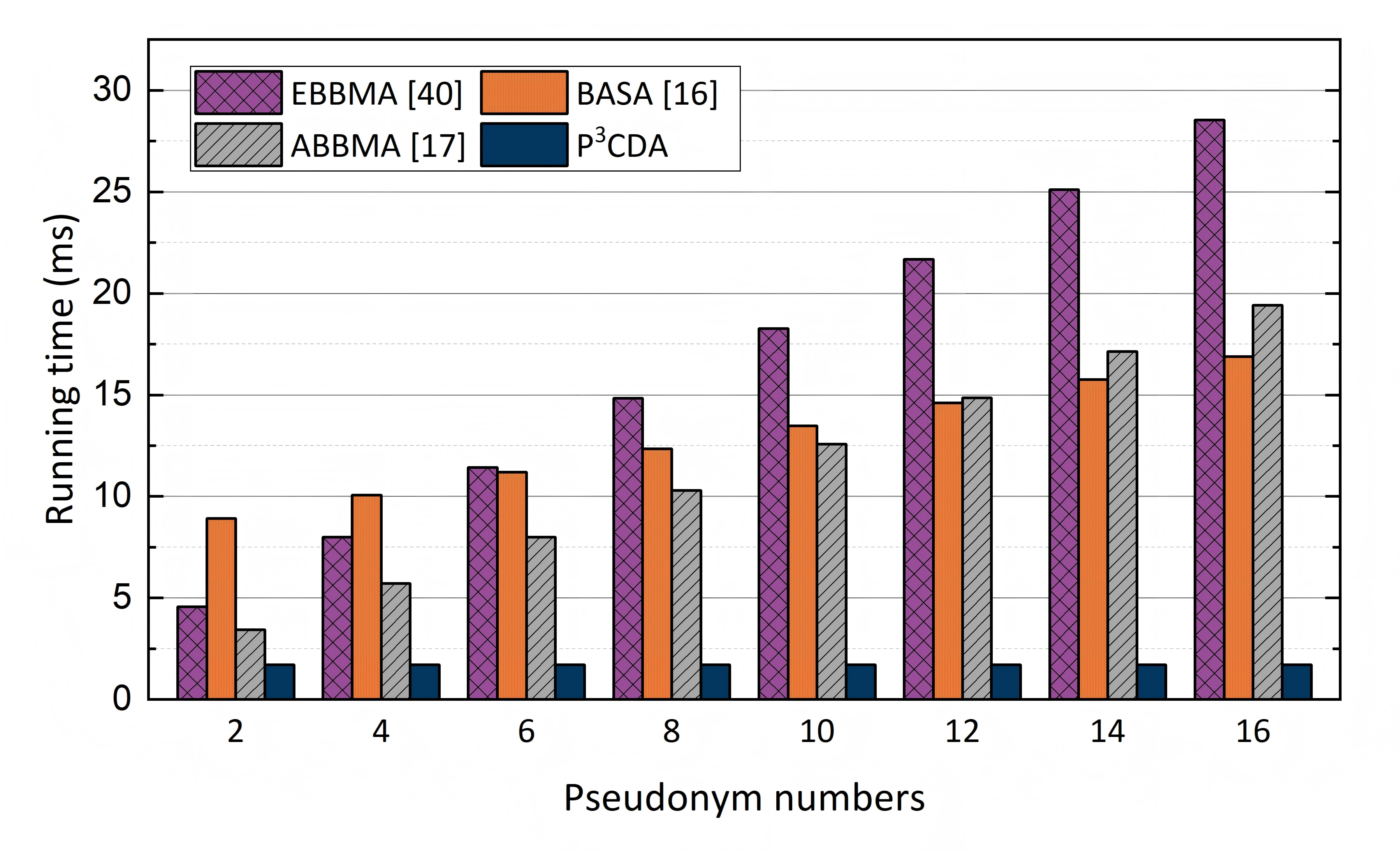}
    \caption{The running time on the TA side}
  \end{subfigure}

  \caption{The running time of pseudonym-based CDA schemes during the registration process.}
  \label{fig:ecdsa_comparison}
\end{figure}

\begin{figure}[t]
  \centering
  \begin{subfigure}[b]{0.24\textwidth}
    \includegraphics[width=\textwidth]{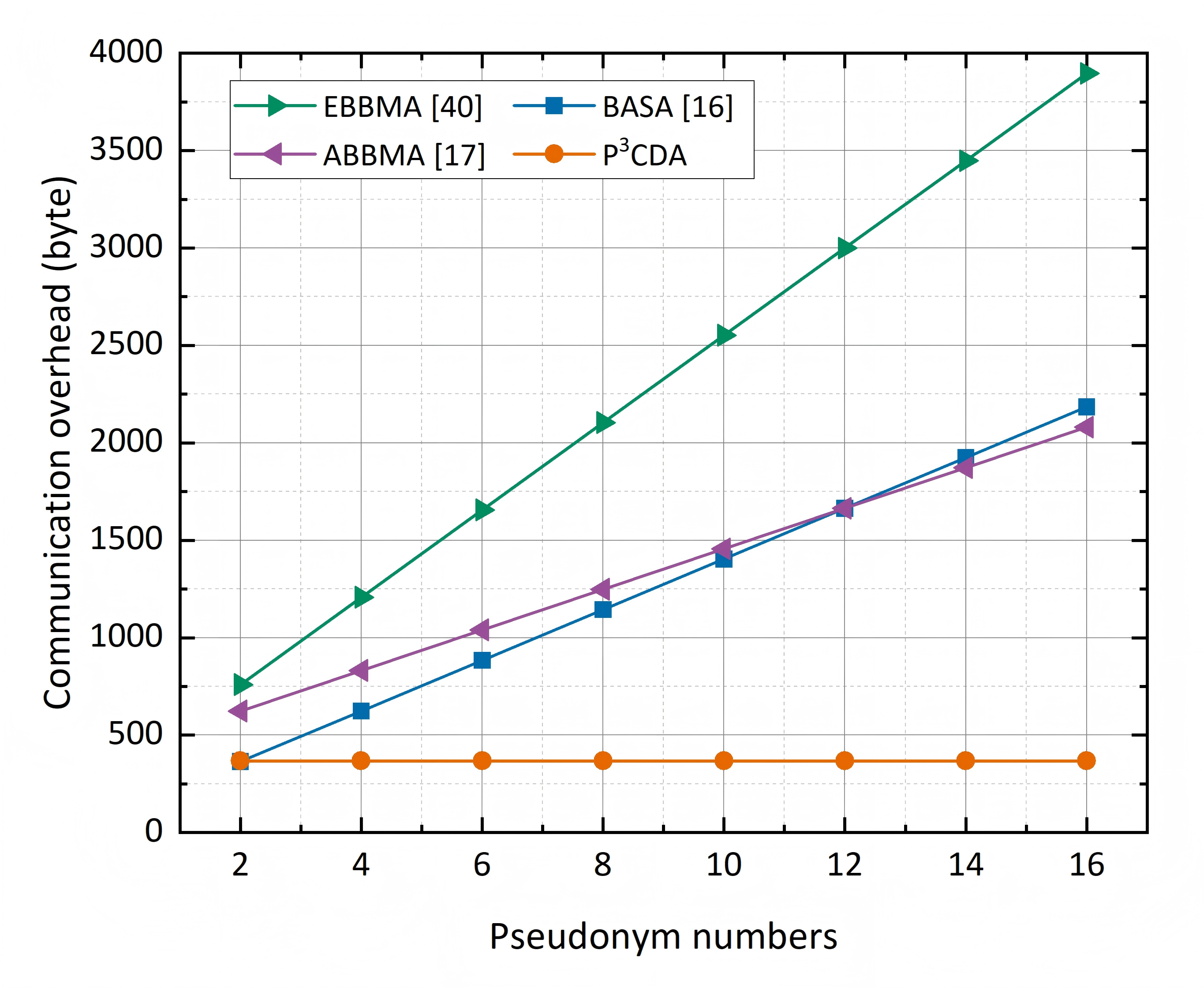}
    \caption{Comm. overhead}
  \end{subfigure}
  \hfill
  \begin{subfigure}[b]{0.24\textwidth}
    \includegraphics[width=\textwidth]{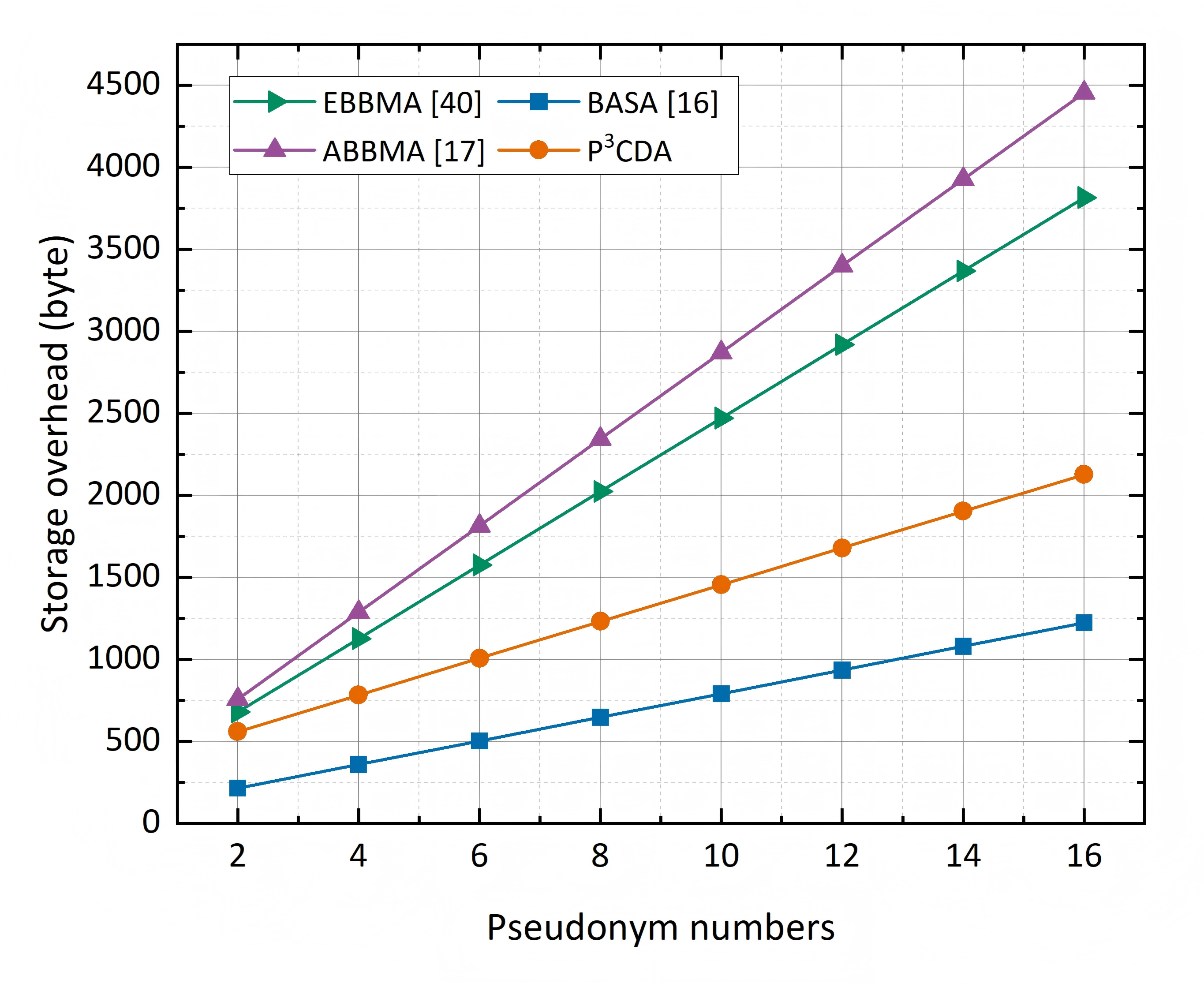}
    \caption{Drone stor. overhead}
  \end{subfigure}

  \vskip\baselineskip
  \begin{subfigure}[b]{0.24\textwidth}
    \includegraphics[width=\textwidth]{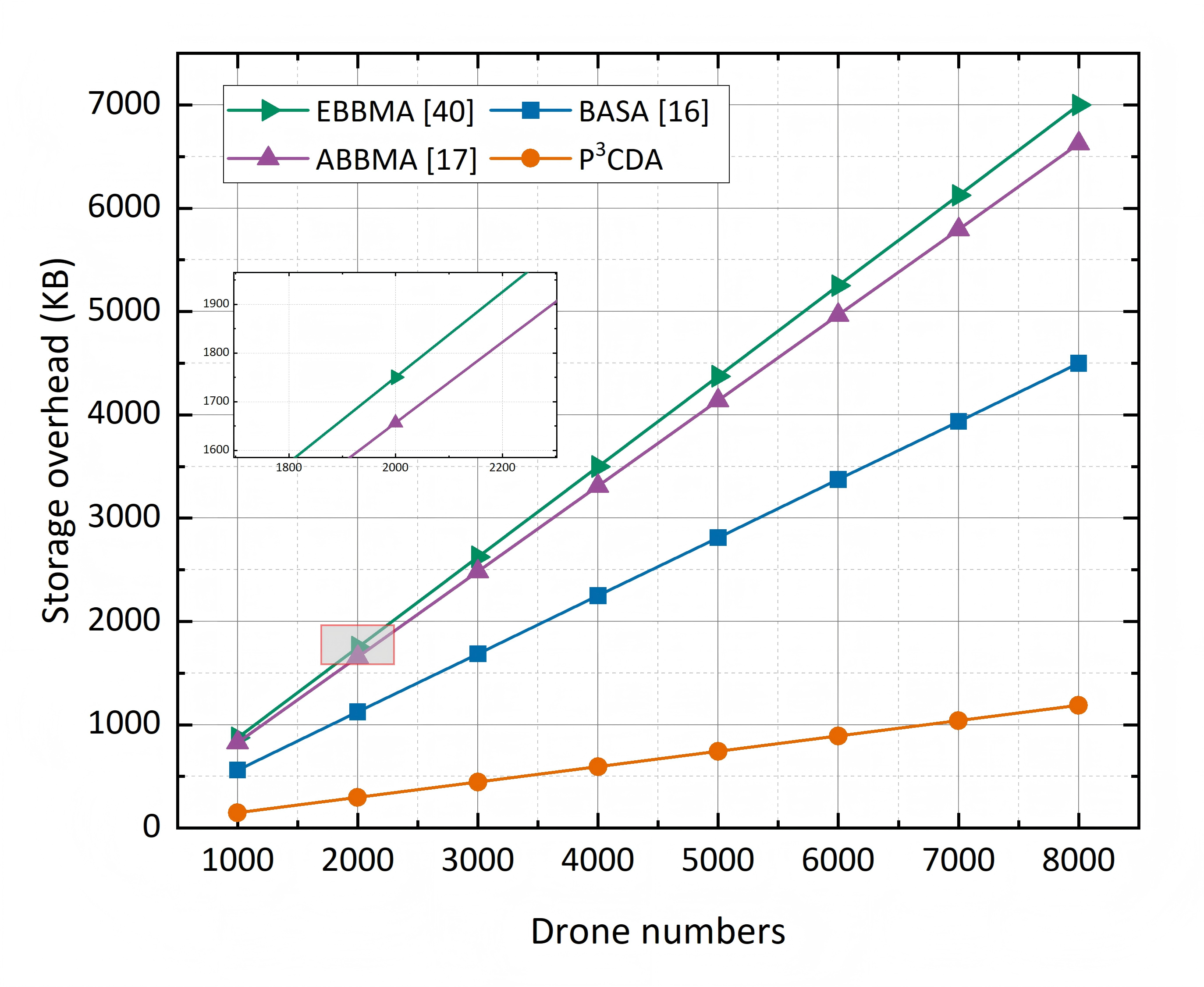}
    \caption{TA stor. overhead ($n=8$)}
  \end{subfigure}
  \hfill
  \begin{subfigure}[b]{0.24\textwidth}
    \includegraphics[width=\textwidth]{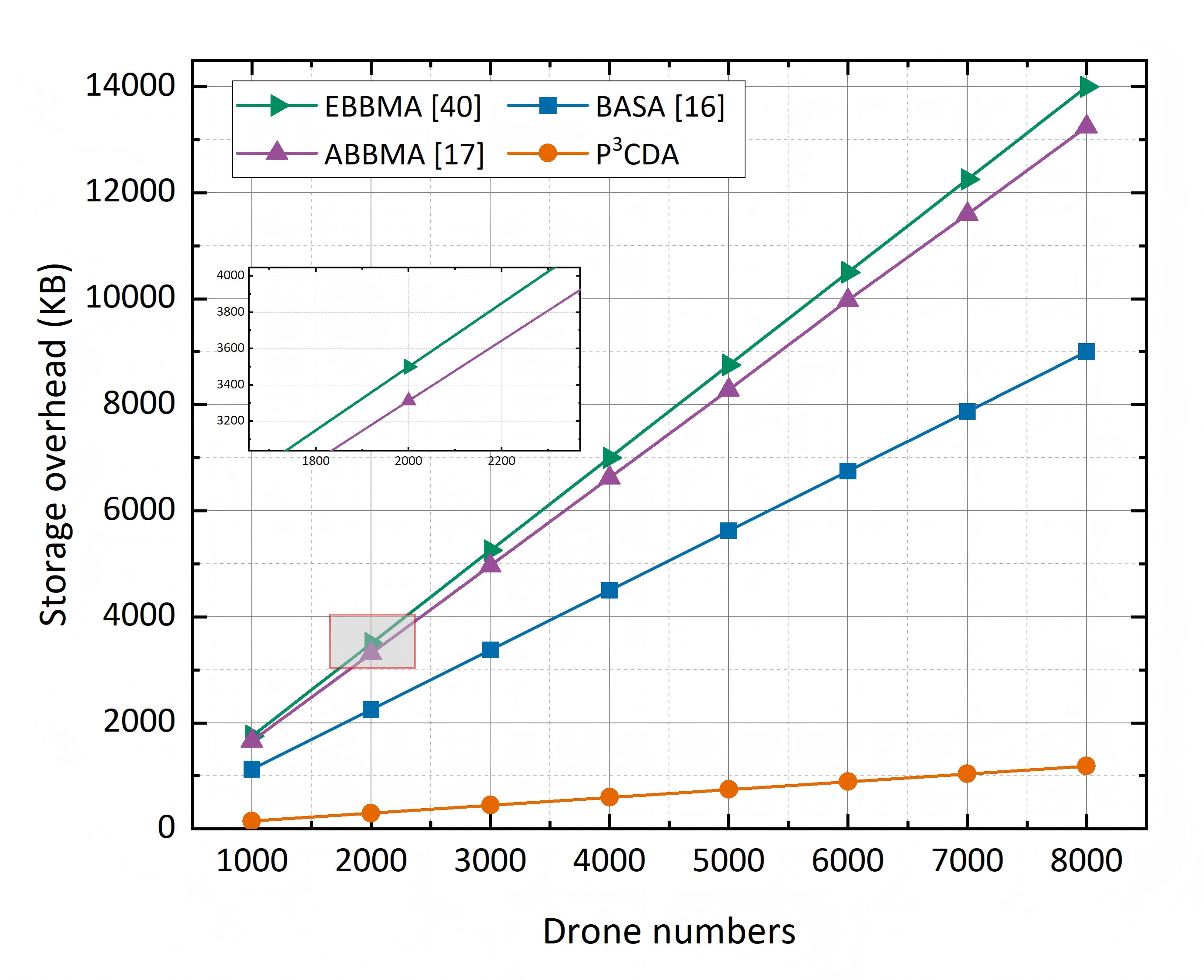}
    \caption{TA stor. overhead ($n=16$)}
  \end{subfigure}

  \caption{Evaluation of communication and storage overhead in pseudonym based CDA schemes during the registration process.}
  \label{fig:ecdsa_comparison}
\end{figure}

\subsection{Simulation Results}
 Fig. 7 presents the running times of different pseudonym-based CDA schemes during the registration phase. In P\textsuperscript{3}CDA, the drone performs local pseudonym generation with a computational cost of $(n+3) T_{m}+(n+1) T_{h}$, and the generation and registration of 16 pseudonyms take approximately 76.24 ms. In comparison, BASA and EBBMA \cite{b42} achieve lower registration latency because the TA directly generates and distributes key material for each device. However, this centralized distribution poses significant security risks, as a compromised TA could result in the disclosure of all registered drones' private keys. Moreover, it places a heavy computational burden on the TA, limiting its scalability for large-scale IoD scenarios. On the TA side, P\textsuperscript{3}CDA combines Merkle hash trees with IBC to efficiently register device identities and batch pseudonyms. The computational overhead is only  $3T_{m}+2T_{h}$, roughly 1.71 ms. This cost is independent of the number of pseudonyms, giving P\textsuperscript{3}CDA the lowest TA-side running time among all schemes while ensuring low latency and scalability.

Fig. 8(a) illustrates the communication overhead of pseudonym-based CDA schemes during the registration process. In P\textsuperscript{3}CDA, since each drone constructs the pseudonym MHT, only the root value needs to be uploaded in the registration message to complete identity binding, resulting in a fixed overhead. The communication overhead is $3|\mathbb{G}|+5|p|+|ID|+|T|$, with a total size of approximately 368 bytes. In contrast, other schemes must contain pseudonym information in the registration message, leading to communication overhead that grows linearly with the number of pseudonyms. In particular, EBBMA and ABBMA incur overheads of $(3n+4)|\mathbb{G}|+(n+1)|p|+3|T|$ and $(n+5)|\mathbb{G}|+(n+2)|p|+(n+2)|T|+|ID|$, respectively. When performing a registration process with 16 pseudonyms, the corresponding communication costs amount to approximately 3,896 bytes and 2,080 bytes, respectively.

Fig. 8(b)–8(d) illustrate the storage overhead of pseudonym-based CDA schemes during the registration phase. On the drone side, P\textsuperscript{3}CDA generates pseudonyms and constructs the MHT, which incurs certain storage costs. Its storage overhead is $(n+4)|\mathbb{G}| + (n+2)|p| + (n+1)|ID| + (n+1)|T|$, corresponding to about 2,128 bytes for storing 16 pseudonyms. This overhead is lower than that of EBBMA (3,186 bytes) and ABBMA (4,456 bytes), but slightly larger than BASA. However, BASA depends on centralized pseudonym distribution and requires secure channels to transmit substantial private key material. On the TA side, P\textsuperscript{3}CDA achieves high storage efficiency, with an overhead of $|\mathbb{G}| + 2|p| + 2|ID| + |T|$. Processing pseudonym information for 1,000 drones requires about 1,188 KB, and this overhead remains constant regardless of the number of pseudonyms requested per device. In contrast, other schemes rely on centralized pseudonym distribution. To ensure the traceability and correctness of pseudonyms, the TA must store substantial data that grows linearly with the number of pseudonyms requested by each device. For example, in ABBMA, 1,000 drones each applying for 16 pseudonyms would require the TA to maintain about 13,250 KB of storage, posing scalability concerns for large-scale IoD scenarios.

\begin{figure}[t]
  \centering
  \begin{subfigure}[b]{0.49\textwidth}
    \includegraphics[width=\textwidth]{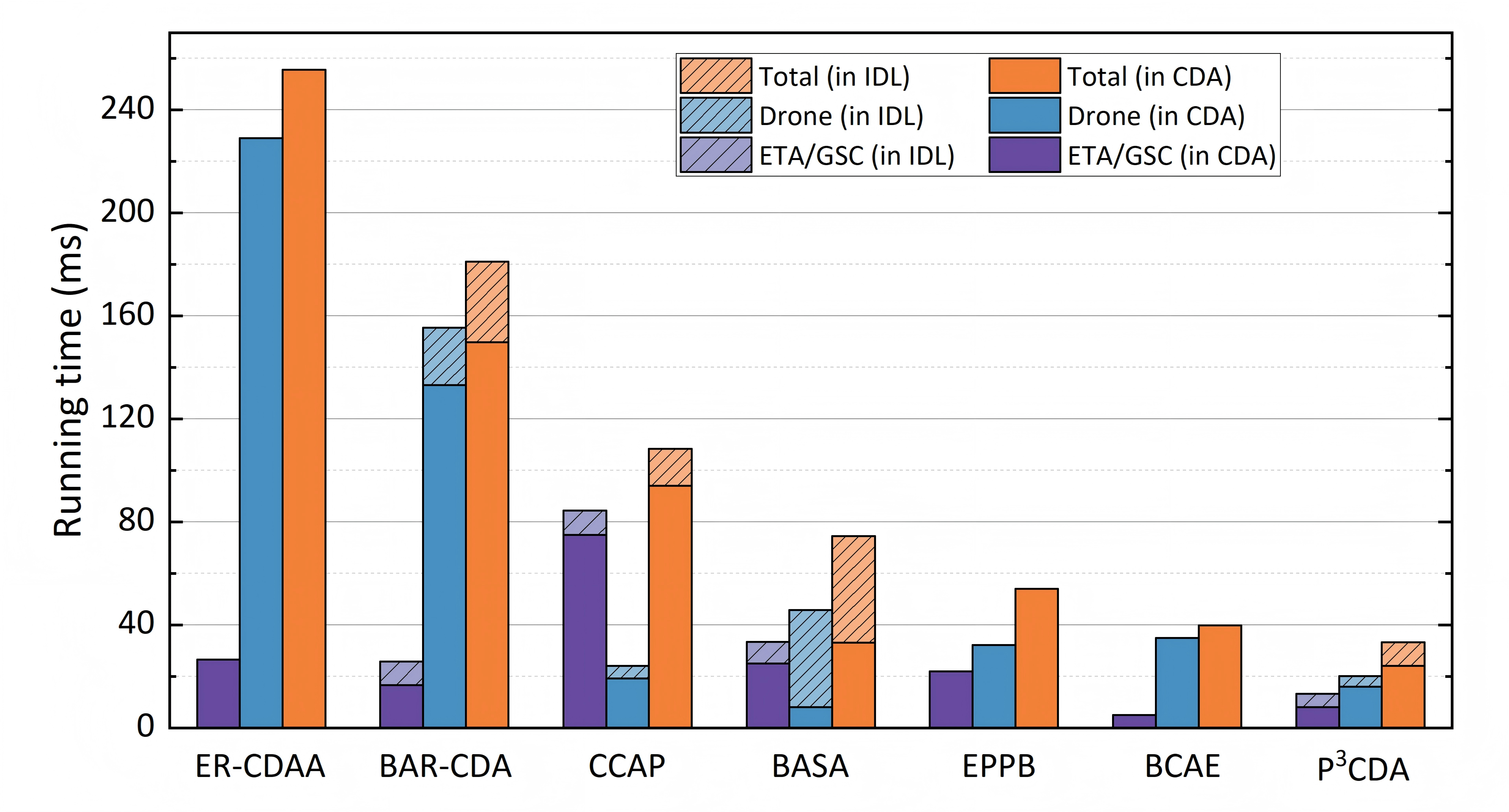}
    \caption{Computational cost}
  \end{subfigure}
  \hfill
  \begin{subfigure}[b]{0.49\textwidth}
    \includegraphics[width=\textwidth]{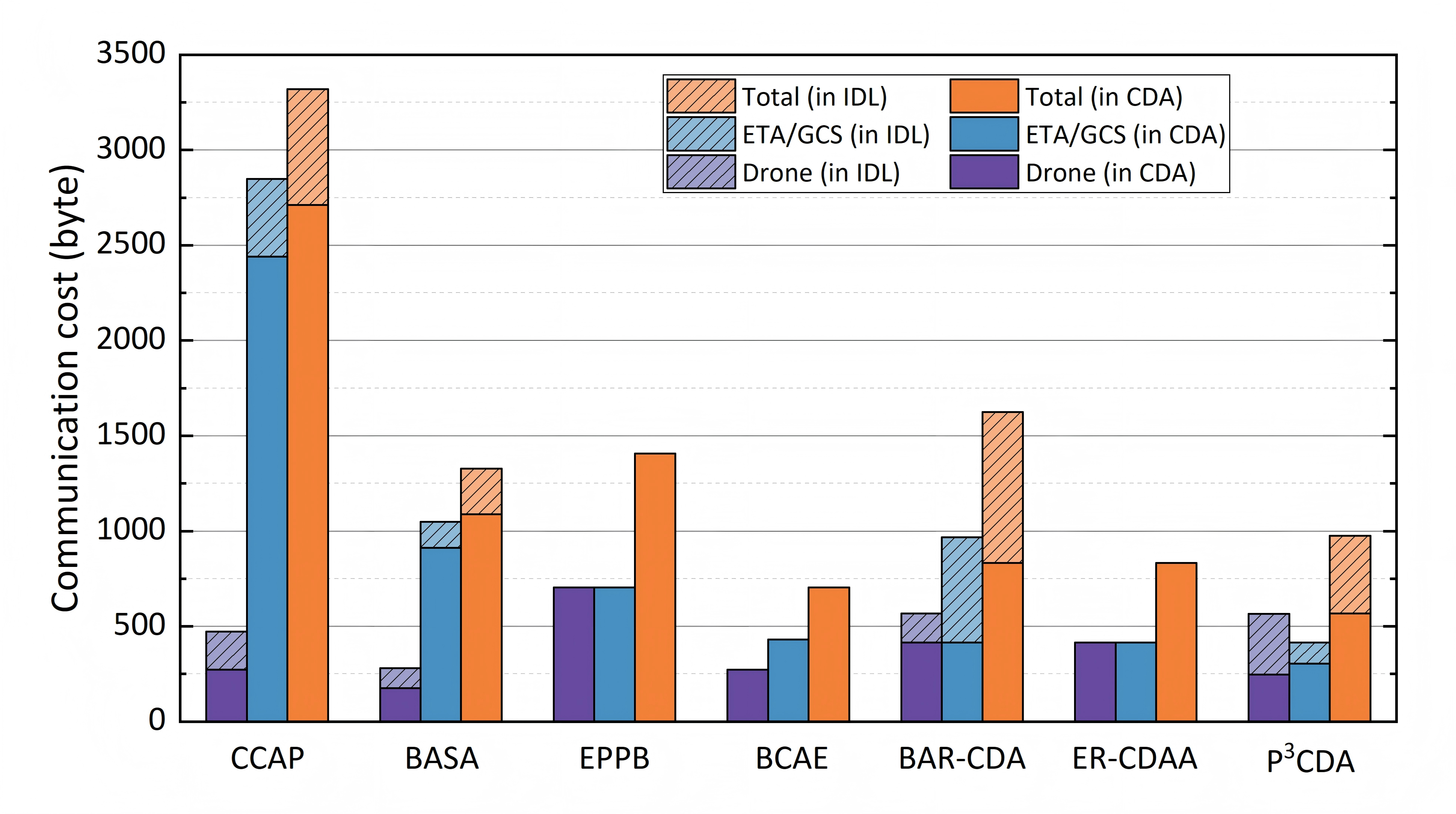}
    \caption{Communication cost}
  \end{subfigure}

  \caption{Simulation results of different CDA schemes during the authentication process.}
  \label{fig:ecdsa_comparison}
\end{figure}

Furthermore, we implemented all CDA schemes analyzed in the previous subsection and evaluated their execution time and communication overhead in completing mutual authentication. The results are presented in Fig. 9 (a) and (b). It can be observed in Fig. 9 (a) that P\textsuperscript{3}CDA achieves the lowest runtime on the drone side, totaling approximately 13.22 ms, with the IDL and CDA processes for around 5.14 and 8.08 ms, respectively. Although P\textsuperscript{3}CDA's total runtime is slightly higher than that of BCAE, BCAE lacks support for any identity privacy protection. Meanwhile, as discussed in the previous subsection, the total runtimes of the remaining comparison schemes are significantly higher than that of P\textsuperscript{3}CDA, further demonstrating that our scheme maintains computational efficiency while preserving identity privacy. 

In terms of communication overhead, as shown in Fig. 9 (b), P\textsuperscript{3}CDA incurs 550 and 432 bytes of data transmission during the IDL and CDA processes, respectively. The overall transmission overhead of P\textsuperscript{3}CDA is slightly higher than the 832 bytes required by ER-CDAA, as it includes additional Merkle path proof in the transmitted parameters. However, this overhead remains acceptable. On one hand, P\textsuperscript{3}CDA significantly outperforms ER-CDAA in terms of computational efficiency. On the other hand, as discussed earlier, P\textsuperscript{3}CDA supports an efficient batch pseudonym login approach, which helps amortize the average communication cost during the IDL phase and improves overall communication performance.

\begin{table}[ht]
\caption{Evaluation of the Pseudonym Updating Process}
\centering
\renewcommand{\arraystretch}{1.2}
\begin{tabular}{c|cc|c} 
\specialrule{1.1pt}{1.1pt}{1.1pt} 
\multirow{2}{*}{Methods} &
\multicolumn{2}{c|}{Computational Overhead} &
\multirow{2}{*}{Communication Overhead} \\
\cmidrule(lr){2-3}
& Drone & TA & \\
\specialrule{1.1pt}{1.1pt}{1.1pt} 

Method \#1 & 12.03~ms & 0.57~ms & 1664~bits \\
\rowcolor[gray]{0.9} 
\textbf{Method \#2} & -- & 1.71~ms & 1408~bits \\
\specialrule{1.1pt}{1.1pt}{0pt} 
\end{tabular}

\vspace{0.5em}
\parbox{0.48\textwidth}{
\footnotesize
\textbullet\ Method \#1 requires re-registration after pseudonym expiration; Method \#2 supports efficient pseudonym updates using a chameleon hash.\\
}
\end{table}

In order to evaluate the effectiveness of the pseudonym update phase in P\textsuperscript{3}CDA, we designed two comparison methods. Method \#1 serves as the baseline approach, which does not utilize the chameleon hash during either the registration or update phases. As a result, the identity registration process must be re-executed once all pseudonyms have expired. Method \#2 is the pseudonym update approach based on the chameleon hash, as proposed in this paper. The evaluation results are presented in Table V. Method \#1 requires drones to re-execute key generation, such as computing $R_i$ and ${PK}_i$, resulting in a computational overhead of $3T_{m}+T_{h}$, approximately 12.03 ms. In contrast, method \#2 avoids this computational overhead by leveraging the collision property of the chameleon hash, allowing the drone to update its pseudonym without re-generating these parameters. Moreover, since method \#2 does not require re-transmission of these parameters during the update process, its communication overhead is lower than that of method \#1, with 176 bytes compared to 208 bytes. This further demonstrates P\textsuperscript{3}CDA’s advantage in both computational and communication efficiency.

\section{Conclusion}
In this paper, we proposed P\textsuperscript{3}CDA, a secure, efficient, and conditional privacy-preserving cross-domain authentication scheme for IoD. First, P\textsuperscript{3}CDA features an efficient pseudonym management mechanism that enables batch registration, fast verification, and flexible updates through the integration of IBC, MHT, and chameleon hash. This mechanism enhances identity privacy while significantly reducing resource overhead. Second, we designed a secure authentication process that allows drones to perform cross-domain authentication using authorized pseudonyms, while enabling the TA and ETA to trace malicious drones and revoke their pseudonym authorizations, thereby achieving a balance between security and controllability. We verified the security of P\textsuperscript{3}CDA through a game-based proof, ProVerif tool, and heuristic analysis, demonstrating its capability to ensure mutual authentication, secure session key negotiation, and resistance to multiple security attacks. The experimental evaluations further analyzed the efficiency of P\textsuperscript{3}CDA in terms of computational, storage, and communication overhead using both theoretical analysis and simulation evaluation. Future work will focus on developing post-quantum cryptography-based cross-domain authentication schemes to address emerging threats from quantum computing.

\end{document}